\documentclass[aip,preprint,superscriptaddress,longbibliography,floatfix]{revtex4-2}

\usepackage{amsmath,braket}
\usepackage{graphicx}
\usepackage{mathtools}
\usepackage{color}
\usepackage{xcolor}
\usepackage{setspace}
\usepackage[pagewise]{lineno}
\makeatletter
\let\@fnsymbol\@fnsymbol@latex
\@booleanfalse\altaffilletter@sw
\makeatother

\graphicspath{{.}{Figures/}}

\begin{document}
\title{Direct observation of photon bound states using a single artificial atom}

\author{Natasha Tomm}
\email{natasha.tomm@unibas.ch}
\thanks{Natasha Tomm and Sahand Mahmoodian contributed equally to this work.}
\affiliation{Department of Physics, University of Basel, Klingelbergstrasse 82, CH-4056 Basel, Switzerland}

\author{Sahand Mahmoodian}
\email{sahand.mahmoodian@sydney.edu.au}
\affiliation{Centre for Engineered Quantum Systems, School of Physics, The University of Sydney, Sydney, NSW 2006, Australia}
\affiliation{Institute for Theoretical Physics, Institute for Gravitational Physics (Albert Einstein Institute),
Leibniz University Hannover, Appelstra{\ss}e 2, 30167 Hannover, Germany}

\author{Nadia O. Antoniadis}
\affiliation{Department of Physics, University of Basel, Klingelbergstrasse 82, CH-4056 Basel, Switzerland}

\author{R\"{u}diger Schott}
\affiliation{Lehrstuhl f\"{u}r Angewandte Festk\"{o}rperphysik, Ruhr-Universit\"{a}t Bochum, D-44780 Bochum, Germany}

\author{Sascha R.\ Valentin}
\affiliation{Lehrstuhl f\"{u}r Angewandte Festk\"{o}rperphysik, Ruhr-Universit\"{a}t Bochum, D-44780 Bochum, Germany}

\author{Andreas D.\ Wieck}
\affiliation{Lehrstuhl f\"{u}r Angewandte Festk\"{o}rperphysik, Ruhr-Universit\"{a}t Bochum, D-44780 Bochum, Germany}

\author{Arne Ludwig}
\affiliation{Lehrstuhl f\"{u}r Angewandte Festk\"{o}rperphysik, Ruhr-Universit\"{a}t Bochum, D-44780 Bochum, Germany}

\author{Alisa Javadi}
\affiliation{Department of Physics, University of Basel, Klingelbergstrasse 82, CH-4056 Basel, Switzerland}

\author{Richard J.\ Warburton}
\affiliation{Department of Physics, University of Basel, Klingelbergstrasse 82, CH-4056 Basel, Switzerland}

\date{\today}

\begin{abstract}
{\bf The interaction between photons and a single two-level atom constitutes a fundamental paradigm in quantum physics. The nonlinearity provided by the atom means that the light-matter interaction depends strongly on the number of photons interacting with the two-level system within its emission lifetime. This nonlinearity results in the unveiling of strongly correlated quasi-particles known as photon bound states, giving rise to key physical processes such as stimulated emission and soliton propagation. While signatures consistent with the existence of photon bound states have been measured in strongly interacting Rydberg gases, their hallmark excitation-number-dependent dispersion and propagation velocity have not yet been observed. Here, we report the direct observation of a photon-number-dependent time delay in the scattering off a single semiconductor quantum dot coupled to an optical cavity. By scattering a weak coherent pulse off the cavity-QED system and measuring the time-dependent output power and correlation functions, we show that single photons, and two- and three-photon bound states incur different time delays of 144.02\,ps, 66.45\,ps and 45.51\,ps respectively. The reduced time delay of the two-photon bound state is a fingerprint of the celebrated example of stimulated emission, where the arrival of two photons within the lifetime of an emitter causes one photon to stimulate the emission of the other from the atom.}
\end{abstract}

\maketitle
\normalsize

Photons do not easily interact with one another. This property is commonly exploited to communicate over long distances using optical fibres. Interaction between photons is desired however for classical and quantum information processing, but requires a highly nonlinear medium. Optical nonlinear processes are employed in a range of photonic applications such as frequency conversion, optical modulation, light amplification, and sensing\,\cite{BoydBook2020, Rolston&Phillips2002, Chang2014}. In the limit where the optical nonlinearity is significant on the scale of a few photons, one can observe quantum nonlinear phenomena, for instance via the optical correlation functions \cite{Chang2014, Chang2018RMP}. One manifestation of the nonlinearity is the presence of two and higher-order photon bound states. Photons in these bound states are strongly correlated such that the likelihood of observing a photon at any one time is fixed, but once one photon is detected, the arrival of another is much more likely than at a random time. We emphasize that photon bound states are distinct from bunched photon states as photon bound states are quasiparticles that have their own dispersion relation and are eigenstates of the underlying Hamiltonian that describes the nonlinear medium. It has recently been predicted theoretically that the photon-number-dependent propagation velocity of photon bound states can lead to the formation of highly-entangled, ordered states of light \cite{Mahmoodian2020}. Photon bound states have been predicted to exist in a number of systems such as unidirectional waveguide quantum electrodynamics (QED) \cite{Rupasov1982JETP, Rupasov1984JETP, Shen2007PRA} and strongly correlated Rydberg gases \cite{Maghrebi2015PRL}. In the latter case, experimental observations consistent with their presence have been reported \cite{Firstenberg2013, Stiesdal2018PRL, Liang2018}. A direct observation is however lacking. To observe directly the photon bound states we examine the propagation of few-photon wavepackets strongly interacting with a single atom, in practice a semiconductor quantum dot (QD) coupled to a one-sided cavity.

The experimental setup is schematically depicted in Fig.~1a: Gaussian pulses of light are guided via a circulator to the one-sided QD-cavity system. The light is back-scattered and redirected by the circulator towards a Hanbury Brown-Twiss (HBT) setup, with single-photon detectors that record the time of arrival $\tau$ of individual photons. By launching a weak coherent pulse with average photon number $\bar n \ll 1$, one can probe directly the scattering dynamics of single-photon pulses via power measurements $P(\tau) = G^{(1)}(\tau)$. Conversely, the second-order correlation function $G^{(2)}(\tau_{ch1},\tau_{ch2})$ is insensitive to the single-photon Fock component and is used to study two-photon scattering dynamics. We also measure the third-order correlation function $G^{(3)}(\tau_{ch1},\tau_{ch2},\tau_{ch3})$ to probe the dynamics of the three-photon component. For higher-order correlation functions, we can determine the equal-time correlators, i.e.\ when $\tau_{ch1}=\tau_{ch2}=...=\tau_{chn}$. For the coherent state the equal-time correlator $G^{(n)}(\tau, \tau, \ldots, \tau) = P(\tau)^n$. Deviations from this indicate that the photons undergo a nonlinear scattering process.


Single-photon states and two-photon states undergo distinct dynamics when scattering off the cavity-QED system. We can understand these dynamics by examining the one- and two-photon scattering eigenstates. In our cavity-QED setup the QD couples almost perfectly to the cavity, which in turn has negligible undesired losses and we thus model our system as being lossless. The single-photon scattering eigenstates are plane waves that are transmitted through the system with a transmission coefficient\cite{Auffeves-Garnier2007PRA, Shi2011PRA, Rephaeli2012IEEE}
\begin{equation}
t_k = \frac{\Delta_{\rm L} - \Delta_{\rm C} - g^2/\Delta_{\rm L} - i \kappa/2}{\Delta_{\rm L} - \Delta_C - g^2/\Delta_{\rm L} + i \kappa/2},
\end{equation}
\noindent where $\Delta_{\rm L}=\omega_{\rm L}-\omega_{\rm QD}$ is the angular frequency detuning of the photon and the QD, $\Delta_{\rm C}=\omega_{\rm C}-\omega_{\rm QD}$ the detuning between the cavity resonance and the QD, $g$ the atom-cavity coupling, and $\kappa$ the cavity loss rate. Under the lossless assumption the scattering amplitude is unitary, $|t_k|=1$, but scattering imparts a frequency-dependent phase on the photon. The importance of the frequency-dependent phase is highlighted when scattering Gaussian pulses off the cavity-QED system. Defining $e^{i \phi(k)} = t_k$ we then have $\phi(k) = -i \ln{(t_k)}$. As in standard Gaussian pulse propagation \cite{SalehTeichBook}, the first to third derivatives of $\phi(k)$ then give the delay $\Delta\tau_1(\Delta_{\rm L},\Delta_{\rm C})$, broadening and chirp, and distortion $d_1(\Delta_{\rm L},\Delta_{\rm C})$ of the Gaussian pulse upon scattering off the quantum system, respectively. 
On resonance ($\Delta_{\rm L}=0,\Delta_{\rm C}=0$), the delay is $\Delta\tau_1(0,0) = 4/\Gamma$, where $\Gamma = 4g^2/\kappa$ is the Purcell-enhanced decay rate. The distortion is given by $d_1(0,0) = -32(1 - 3 \Gamma/\kappa)/\Gamma^3$.

The physics of two-photon scattering is richer as the energy of the individual photons is not necessarily conserved, which leads to photon correlations. The two-photon scattering matrix has previously been computed \cite{Shi2011PRA, Rephaeli2012}, but here, we diagonalise the scattering matrix and show that the two-photon eigenstates contain a subspace of two-photon bound states (photonic dimers). The two photons composing the bound state are entangled with each other. In the relative two-photon coordinate $x=x_1-x_2$, the photons are exponentially localised; however, since the two-photon energy is conserved, they take the form of a plane wave $e^{i E x_c}$ in the two-photon centre-of-mass coordinate $x_c = (x_1+x_2)/2$ with a common two-photon frequency $E$. The exponential localisation of these states has a width of approximately $1/\Gamma$ in the difference coordinate, which means that the states are formed by stimulated emission of the cavity-QED system as the two photons interact with the system within its relaxation time. This distinctly correlated interaction between the photons leads to this eigenstate having its own distinct transmission coefficient $t_E^{(2)}$ and dispersion in comparison to the single-photon eigenstate and therefore undergoes different delays, broadening and distortion. We compute the general form of $t_E^{(2)}$ numerically, but in the limit of a broadband cavity ($\kappa \gg \Gamma$), from perturbation theory we find that the delay of the two-photon bound state in the centre-of-mass coordinate is $\Delta\tau_2(0,0) = 1/\Gamma + 3/\kappa$. Due to the influence of stimulated emission, the bound state therefore undergoes a reduced delay. The distortion is $d_2(0,0) = -(1 - 3 \Gamma/\kappa)/(2 \Gamma^3)$. In comparison to the single-photon state it thus undergoes a factor of $64$ less distortion. The reduction in distortion was shown to be related to soliton propagation in waveguide QED \cite{Mahmoodian2020}.

In order to fulfil experimentally the two criteria to study the photon-number-dependent scattering dynamics, namely unidirectional light propagation and a strong atom-photon interaction, we employ not a real atom but an artificial atom, a single QD. The QD is embedded in a Fabry-Perot microcavity. The epitaxially grown InAs QDs are part of a semiconductor heterostructure comprising an n-i-p diode and a GaAs/AlAs Bragg reflector, the ``bottom mirror''. The ``top mirror'' consists of a concave, dielectric Bragg mirror fabricated into a silica substrate. The reflectivity of the bottom mirror is significantly higher than that of the top mirror. With the aid of xzy-nanopositioners, one can position a QD in the sample relative to the cavity mode and one can tune the resonance frequency of the cavity to that of the QD's emission. Essential for the unidirectionality condition, the cavity should have only one port. In this system, undesired losses (losses via the bottom mirror, absorption and scattering losses\cite{Najer2021}) account for $\kappa_{\rm loss}/(2\pi)=(0.72\pm0.07)$\,GHz\cite{Tomm2021}, while the total cavity linewidth is $\kappa/(2\pi)=(20.1\pm1.5)$\,GHz, indicating that $\sim96\%$ of the light is back-reflected via the one port of the microcavity, namely the top mirror. The QDs in this sample present a close-to-transform-limited linewidth $\gamma/(2\pi)=0.30$\,GHz. When QD and cavity are coupled, the Purcell-enhanced QD linewidth $\Gamma=F_{\rm P}\cdot\gamma$ becomes $\Gamma/(2\pi)=4.24$\,GHz, where $F_{\rm P}=14.1$ is the Purcell factor. The lifetime of the emitter becomes $\tau_{\rm QD}=37.5$\,ps and the QD-cavity coupling rate $g/(2\pi)=4.62$\,GHz. The strong atom-photon interaction and near-lossless operation result in a near-unitary probability of emission from the QD into the cavity mode, $\beta=F_{\rm P}/(F_{\rm P}+1)=0.93$. Further experimental details are provided in the Supplementary Information.

The direct observation of photon-number-dependent scattering dynamics is presented in Fig.~1b. A weak, coherent Gaussian pulse of temporal FWHM=135\,ps -- about twice the lifetime of the QD, $\sigma\Gamma=2.2$ -- is launched into the input of the optical system. Without any interaction with the cavity-QED system, it propagates through the optical system and arrives at the single-photon detectors at time $\tau=0$ (the pulse peak is represented by the grey dashed line). Upon resonant interaction with the broadband cavity, but in the absence of the quantum emitter, the Gaussian pulse undergoes a shape-preserving and linear transmission, but is delayed by $\Delta\tau_{\rm C}=(29.2\pm0.4)$\,ps (the centre of the Gaussian pulse is represented by the black dashed line). This delay matches the predicted delay for a one-sided cavity $\Delta\tau_{\rm C}=4/\kappa=(30.9\pm2.2)$\,ps (see Supplementary Information). The delay imparted via the elastic scattering of a wavepacket by a resonator is often referred to as a `Wigner delay'\cite{Wigner1955}. In the presence of the QD, i.e.\ when the QD is tuned into resonance with the cavity, we observe that the scattered $n$-photon pulse reveals an $n$-dependent delay, a \textit{quantum Wigner delay}. We inspect the dynamics via the $n^{\rm th}$ equal-time correlators: the single-photon scattering is given by a power measurement, $G^{(1)}(\tau)$, presented as red dots in Fig.~1b, and shows how the output pulse is delayed relative to the input pulse, a result previously observed also for other quantum systems\cite{Bourgain2013,Leong2016,Strauss2019}. The scattered pulse is non-Gaussian in shape. This distortion causes the peak of the pulse to be delayed by a value different from $\Delta\tau_1=4/\Gamma$. The distortion arises from the fact that the spectral components of the pulse probe a sizeable fraction of the components making up the resonance of the cavity-QED system. The observed dynamics in $G^{(1)}(\tau)$ are well captured by the theoretical model (red solid line). The scattering of two-photon states is examined via $G^{(2)}(\tau,\tau)$, Fig.~1b blue dots (theory, solid blue line). We find experimentally that both the delay and distortion in $G^{(2)}(\tau,\tau)$ are significantly reduced compared to those of $G^{(1)}(\tau)$. The theory accounts for the measured $G^{(2)}(\tau,\tau)$ very convincingly. This constitutes a clear observation of two-photon bound states. We interrogate also the third-order equal-time correlator, $G^{(3)}(\tau,\tau,\tau)$, given by the green dots (the green solid line is a Gaussian fit). Both the peak delay and temporal width of the $G^{(3)}(\tau,\tau,\tau)$ curve are reduced further, consistent with the observation of three-photon bound states.

The finite spectral width of the pulses probes the frequency- and photon-number-dependent phase imparted on different photon-number states. For a coherent input pulse we examine the delay experienced by an $n$-photon pulse by comparing the delay at the peak of the scattered pulse $\Delta\tau_n\coloneqq\max(G^{(n)}(\tau,\tau,...,\tau))$ (Fig.~1c). Since longer pulses undergo significantly reduced distortion, for $n=1,2$ we extract the peak delay from measurements with ten different pulse widths ($1.3\leq\sigma\Gamma\leq26.0$). The one- and two-photon delays correspond well to the theoretical predictions (red and blue dashed lines respectively). The single-photon wavepackets undergo a delay $\Delta\tau_1=(144.02\pm26.90)$\,ps; two-photon wavepackets undergo a reduced delay $\Delta\tau_2=(66.45\pm5.97)$\,ps. The reduced delay is a consequence of stimulated emission: the first photon excites the atom, the second photon stimulates the emission of the atom, thereby reducing the total time in which the photons interact with the atom. The three-photon delay is $\Delta\tau_3=(45.51\pm0.09)$\,ps, a further reduction. This measurement of the quantum Wigner delay therefore unveils the existence of few-photon bound states in a direct way. Key to success is the strong nonlinear and unidirectional scattering off the single quantum emitter.

We proceed to examine the auto-correlation functions of the scattered pulses. Figure~1d shows the two-photon auto-correlation map $G^{(2)}(\tau_{ch1},\tau_{ch2})$ of the weak Gaussian pulses scattered off the optical cavity, but in absence of the quantum emitter. The linear response of the cavity displaces the two-dimensional Gaussian structure by $\Delta\tau_{\rm C}$ along the equal-time-of-arrival line ($\tau_{ch1}=\tau_{ch2}$) with respect to the Gaussian structure of the non-interacting pulse centred at $(\tau_{ch1},\tau_{ch2})=(0,0)$. As shown experimentally (theoretically) in Fig.~1e(f), when the pulse interacts with the QD-cavity system the correlated counts are drawn towards the diagonal of the $G^{(2)}$-map (white dashed line), and can no longer be described by a linear transformation of the response to the bare cavity. We examine also the three-photon auto-correlation map $G^{(3)}(\tau_{ch1},\tau_{ch2},\tau_{ch3})$ in Fig.~1g, where the volumetric isosurfaces at 0.05, 0.2, 0.5, 0.75 and 0.90 of the normalised counts are shown. The projections on the axes are the cut-through planes in each of the detection channels at time $\tau=0$. As in the $G^{(2)}$ measurements, there is a strong peak along the diagonal revealing highly correlated three-photon states, as well as faint lateral lobes away from the diagonal\,\cite{Chen2020}. Figure~1g displays a cut-through along the plane defined by $\tau_{ch1}=\tau_{ch2}$ and $\tau_{ch3}$ where the clustering of the coincidence counts along the diagonal are prominently revealed. In all likelihood, this manifests the propagation of three-photon bound states (photonic trimers).

Next, we investigate the behaviour of the single- and two-photon scattering dynamics as a function of the central frequency of the photons. Figure~2a,b show, respectively, the experimental and simulated power signal of the scattered pulse (FWHM = 135 ps) as a function of laser detuning from the QD's resonance $\Delta_{\rm L}/(2\pi)$. The red dashed line in the theoretical model shows where the pulse maximum would occur if distortion effects were disregarded. The cavity is slightly detuned from the QD, $\Delta_{\rm C}/(2\pi)=1.0$\,GHz, which induces a slight spectral asymmetry to the one- and two-photon peak delays presented in Fig.~3c (red and blue dots, respectively). We calculate numerically the dispersion of the peak delays $\Delta\tau_1$ and $\Delta\tau_2$ (red and blue solid lines). The results describe the experimental observations very well. Here too, the red dashed line corresponds to the single-photon case neglecting pulse distortion. The results demonstrate that two-photon bound states experience a much reduced distortion, imperceptible in this system.

The interaction of the two-photon wavefunction with the cavity-QED system strongly depends on the Gaussian pulse width. In Fig.~3a we explore this dependence, where we present on the top row the experimental $G^{(2)}$-map for three different pulse widths, FWHM=(81.86, 256.72, 597.83) ps -- equivalently $\sigma\Gamma=(1.3,4.1,9.6)$. We compare the experimental results to the simulated absolute-square of the full two-photon wavefunction (middle row), which contains contributions from both two-photon bound states and extended states. The total contribution of the bound states alone is shown in the bottom row: the appearance of just the diagonals is evidence that that the bound states contribute to the diagonal of the correlation maps; the extended states contribute to the lobes away from the diagonals. Details of the model are elucidated in the Supplementary Information. The nodal line (absence of coincidence events) that occurs between the diagonal (contribution from the bound states) and lobes (contribution from the extended states) occurs due to the different phase the two states obtain after scattering. The phase approaches $\pi$ for the bound state and is $0$ for the extended states. Finally, the total fraction of the scattered wavefunction in the bound-state subspace depends on the overlap of the two-photon input pulse with the two-photon bound states formed in the cavity-QED system. The theory shows that this overlap has a strong dependence on the input Gaussian pulse duration relative to the lifetime of the quantum emitter, and is largest when the input pulse has a duration $\sim 1/\Gamma$, as shown in Fig.~3b.

We demonstrate here the ability to manipulate and identify highly correlated photonic states in time. The results reveal stimulated emission in its most canonical description, a single quantum emitter interacting with single photons\,\cite{Rephaeli2012}. This achievement represents an important landmark in the development of a variety of quantum technologies. Stimulated emission plays a central role for instance in approximate-quantum cloning of photons\,\cite{Lamas-Linares2002}, a key technology for quantum information processing and networking. The strong dependence of the propagated pulse on photon number can be enhanced by cascading such cavity-QED systems and enables a variety of important applications, such as photon sorting, photon-number-resolving detectors and Bell measurements\,\cite{Ralph2015,Witthaut2012,Yang2022}. The revealing of two-photon bound states upon interaction with a single atom is an appealing resource for the realisation of high-fidelity two-qubit photonic gates, such as controlled-phase gates\,\cite{Chen2021}. Furthermore, the systematic generation of photonic dimers paves the way for significant advances in quantum metrology\,\cite{Giovanetti2011NPHOT}, and quantum-enhanced microscopy and lithography\,\cite{Muthukrishnan_2004,Boto2000}.

\noindent {\bf Acknowledgements} We thank Klemens Hammerer for fruitful discussions. N.T., N.O.A., A.J. and R.J.W.\ acknowledge financial support from SNF project 200020\_204069 and NCCR QSIT. A.J.\ acknowledges support from the European Unions Horizon 2020 Research and Innovation Programme under the Marie Sk{\l}odowska-Curie grant agreement No.\ 840453 (HiFig). S.R.V., R.S., A.L.\ and A.D.W.\ acknowledge gratefully support from DFH/UFA CDFA05-06, DFG TRR160, DFG project 383065199, and BMBF Q.Link.X. S.M.\ acknowledges support from the Australian Research Council (ARC) via the Future Fellowship, ``Emergent many-body phenomena in engineered quantum optical systems'', Project No.\ FT200100844 as well as the ARC Centre of Excellence in Engineered Quantum Systems (EQuS), Project No.\ CE17010000. S.M.\ also acknowledges funding from DFG through CRC 1227 DQ-mat, projects A05 and A06, and ``Nieders\"{a}chsisches Vorab'' through the ``Quantum- and Nano-Metrology (QUANOMET)''.

\noindent {\bf Author contributions} N.T., S.M., A.J. and R.J.W.\ designed the research and experiments. N.T.\ carried out the experiments with participation from N.O.A.\ S.M.\ developed the theoretical model and simulations. N.T.\ and S.M.\ analysed the data. R.S., S.R.V., A.D.W.\ and A.L.\ fabricated the semiconductor device. N.T., S.M.\ and R.J.W.\ wrote the paper with input from all authors.\\

\noindent {\bf Competing interests} The authors declare no competing interests.\\

\noindent {\bf Correspondence and requests for materials} should be addressed to S.M.\ or N.T.\\

\bibliographystyle{naturemag_noURL}
\bibliography{WignerDelay_2021_v1}

\newpage

\begin{figure*}[t!]
\centering
\includegraphics[width=\textwidth]{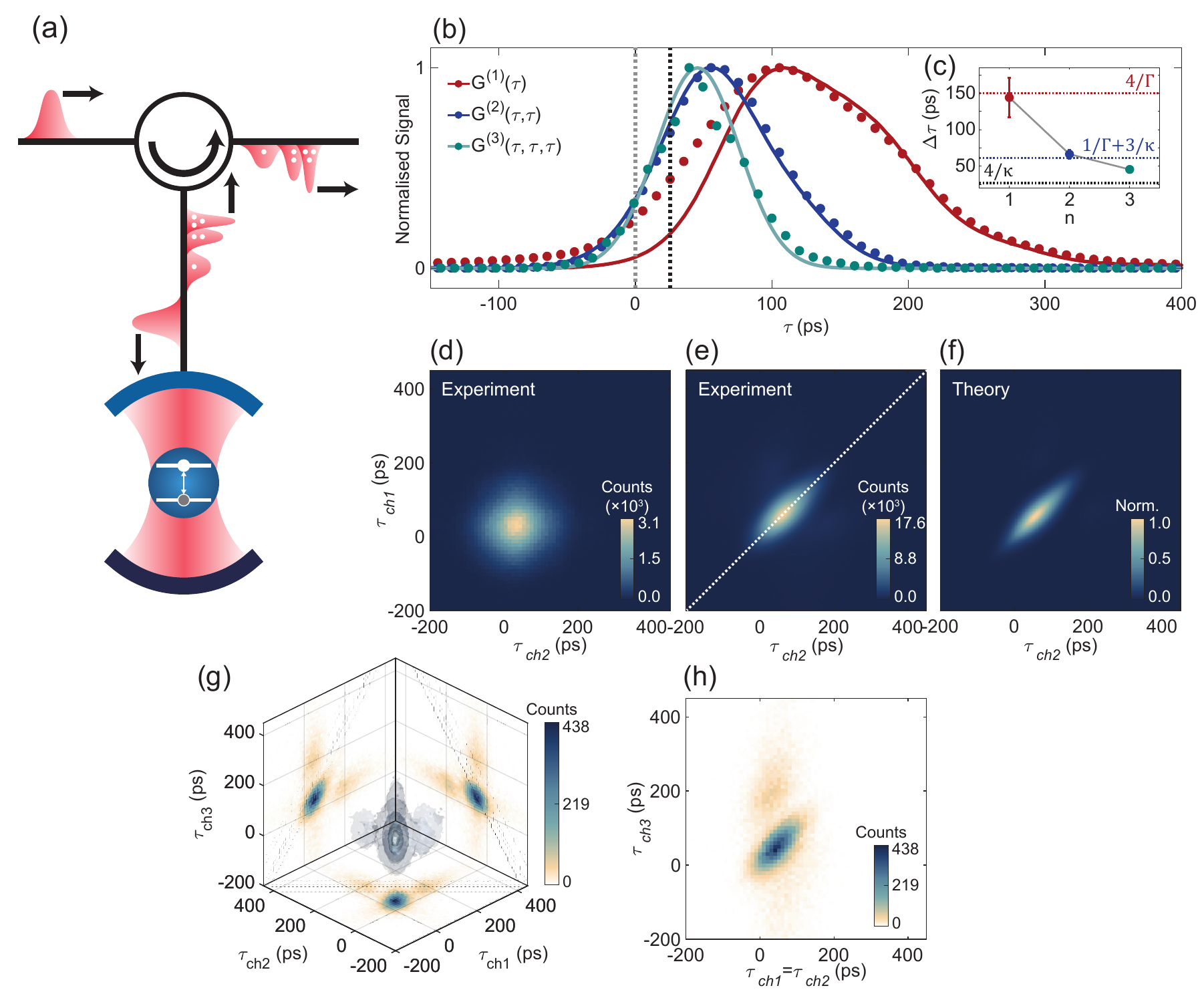}
\caption{Photon-number-dependent pulse scattering. (a) A Gaussian pulse of light is launched into a circulator, which guides the pulse towards a quantum dot (QD), coupled to a one-sided microcavity. Upon interaction with the QD-cavity system, states of light with different photon-number are transported through the system with different time delays. (b) Normalised photon-counts versus delay. For a Gaussian pulse ($\sigma\Gamma=2.2$) launched at time $\tau=0$ (dashed grey line), the propagated pulse undergoes a Wigner delay in the presence of the optical cavity alone (dashed black line). Scattering off the QD-cavity system, the single-photon components $G^{(1)}(\tau)$ -- red points experiment, red solid line theoretical model -- undergo pulse reshaping and arrive with a larger delay than the two-photon bound states $G^{(2)}(\tau,\tau)$ -- blue points experiment, blue solid line theoretical model -- which in turn undergo a larger delay than three-photon states $G^{(3)}(\tau,\tau,\tau)$ -- green points experiment, green solid line Gaussian fit). (c) Average pulse-peak delay $\Delta\tau$ experienced by $n$ photons extracted from low-power resonant measurements for different input pulse widths. (d) Auto-correlation map $G^{(2)}(\tau_{\rm ch1},\tau_{\rm ch2})$ of the pulse following propagation through the entire system in resonance with the optical cavity but in absence of the QD, and (e) in the presence of the QD. The white dashed line represents the equal-time correlation. (f) Simulation of normalised $|\psi_2(\tau_{\rm ch1},\tau_{\rm ch2})|^2$. (g) Auto-correlation map $G^{(3)}(\tau_{\rm ch1},\tau_{\rm ch2},\tau_{\rm ch3})$. The volumes in the 3-dimensional space depict the isosurfaces at 0.05, 0.20, 0.50, 0.75 and 0.90 of the normalised counts, and the projections on each axis are plotted on setting one of the detection times to zero. (h) Cut-through of $G^{(3)}$ at times $\tau_{\rm ch3}$ and $\tau_{\rm ch1}=\tau_{\rm ch2}$.}
\end{figure*}

\begin{figure*}[t!]
\centering
\includegraphics[width=0.5\textwidth]{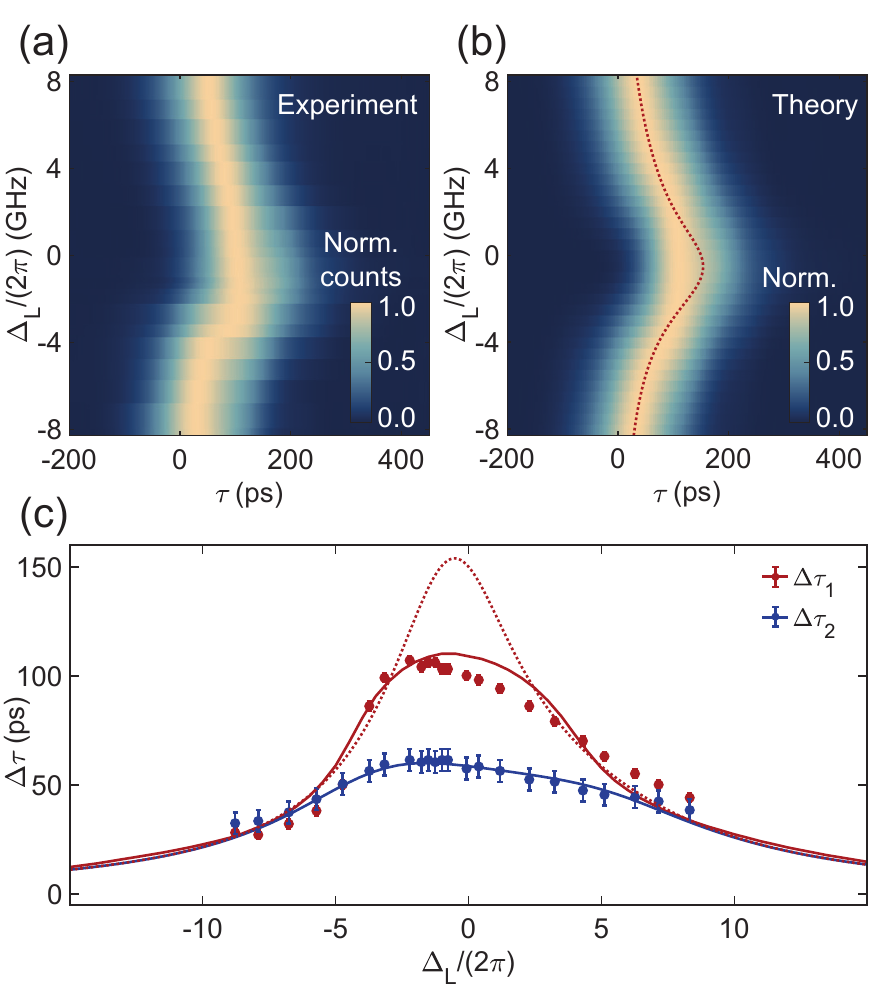}
\caption{Single-photon and two-photon bound state delay dispersion. (a) Experimental $G^{(1)}(\tau)$ as a function of laser detuning $\Delta_{\rm L}/(2\pi)$ for a cavity-QD detuning of $\Delta_{\rm C}/(2\pi)=1.0$\,GHz and for an input pulse with intensity FWHM of $\sim$135\,ps, and (b) respective simulation, where the red dotted line indicates the single-photon component delay in the continuous wave limit. (c) Peak delay $\Delta\tau$ as a function of laser detuning for single photons (red) and two-photon bound states (blue). The solid lines are the numerically simulated peak delays for the single- and two-photon bound states. The red dotted line shows the calculated pulse delay $\Delta\tau_1$ neglecting distortion. In this system, the two-photon bound state propagates without noticeable distortion.}
\end{figure*}

\begin{figure*}[t!]
\centering
\includegraphics[width=0.5\textwidth]{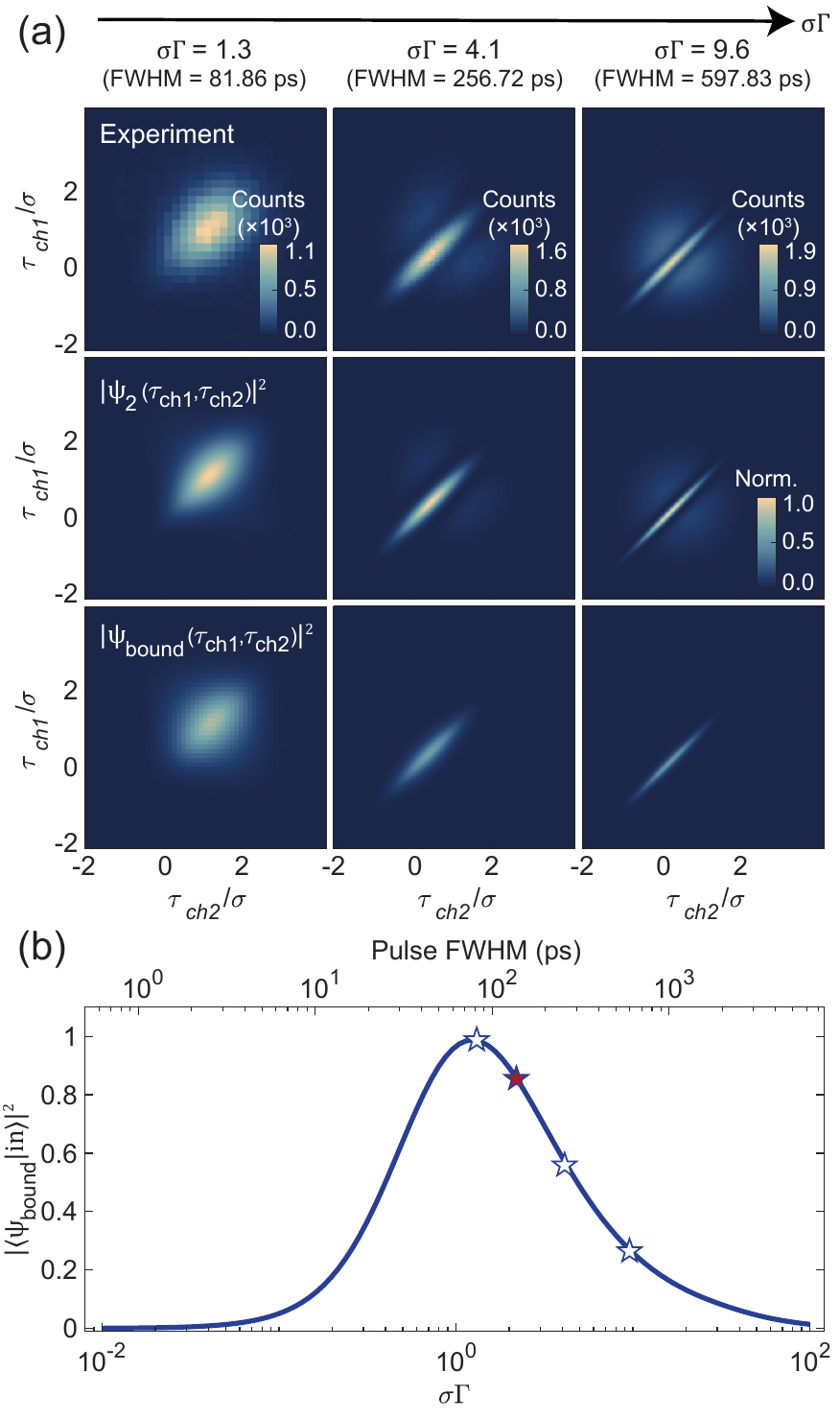}
\caption{Formation of two-photon bound states as a function of input pulse width. (a) Experimental (top row) and simulated (middle row) $G^{(2)}(\tau_{\rm ch1},\tau_{\rm ch2})$, as well as two-photon bound-state contribution (bottom row) for increasing relative pulse widths $\sigma\Gamma = (1.3,4.1,9.6)$ (left to right). The time axes are normalised by the pulse width for better visualisation. (b) Calculated overlap of two-photon bound state and an input state $\ket{in}$ described by a two-photon Gaussian pulse of width $\sigma\Gamma$. The upper x-axis shows the experimental intensity FWHM of the input pulse. The white stars correspond to the pulse widths probed in (a), and the red star the pulse width $\sigma\Gamma=2.2$ showcased in Figures 1 and 2.}
\end{figure*}

\end{document}


\title{Supplementary Information: Direct observation of photon bound states using a single artificial atom}

\author{Natasha Tomm}
\author{Sahand Mahmoodian}
\author{Nadia O. Antoniadis}
\author{R\"{u}diger Schott}
\author{Sascha R.\ Valentin}
\author{Andreas D.\ Wieck}
\author{Arne Ludwig}
\author{Alisa Javadi}
\author{Richard J.\ Warburton}


\maketitle

\section{Microcavity and quantum dot characterisation}
\label{SIsec:cavityQDcharacterisation}

The quantum dot (QD) in a cavity system consists of two main elements. As shown in Supplementary Fig.\,\ref{SIfig:cavQDscans}a, we employ a semiconductor heterostructure consisting of an n-i-p diode with embedded self-assembled InGaAs quantum dots (QDs). This heterostructure is grown on top of a semiconductor distributed Bragg reflector (DBR) that acts as the planar ``bottom'' mirror of the cavity that is nearly 100\% reflective -- the total unwanted losses $\kappa_{\rm loss}/(2\pi)\leq0.72$\,GHz. The ``top'' mirror is made of a fused-silica substrate into which an atomically-smooth crater with radius of curvature $R=12\,\mu$m is machined, onto which a dielectric DBR coating is deposited. Further details of the sample and the system are described in detail in Ref.\,\cite{Tomm2021}, Supplementary Information.

\begin{figure*}[hb!]
\centering
\includegraphics[width=\textwidth]{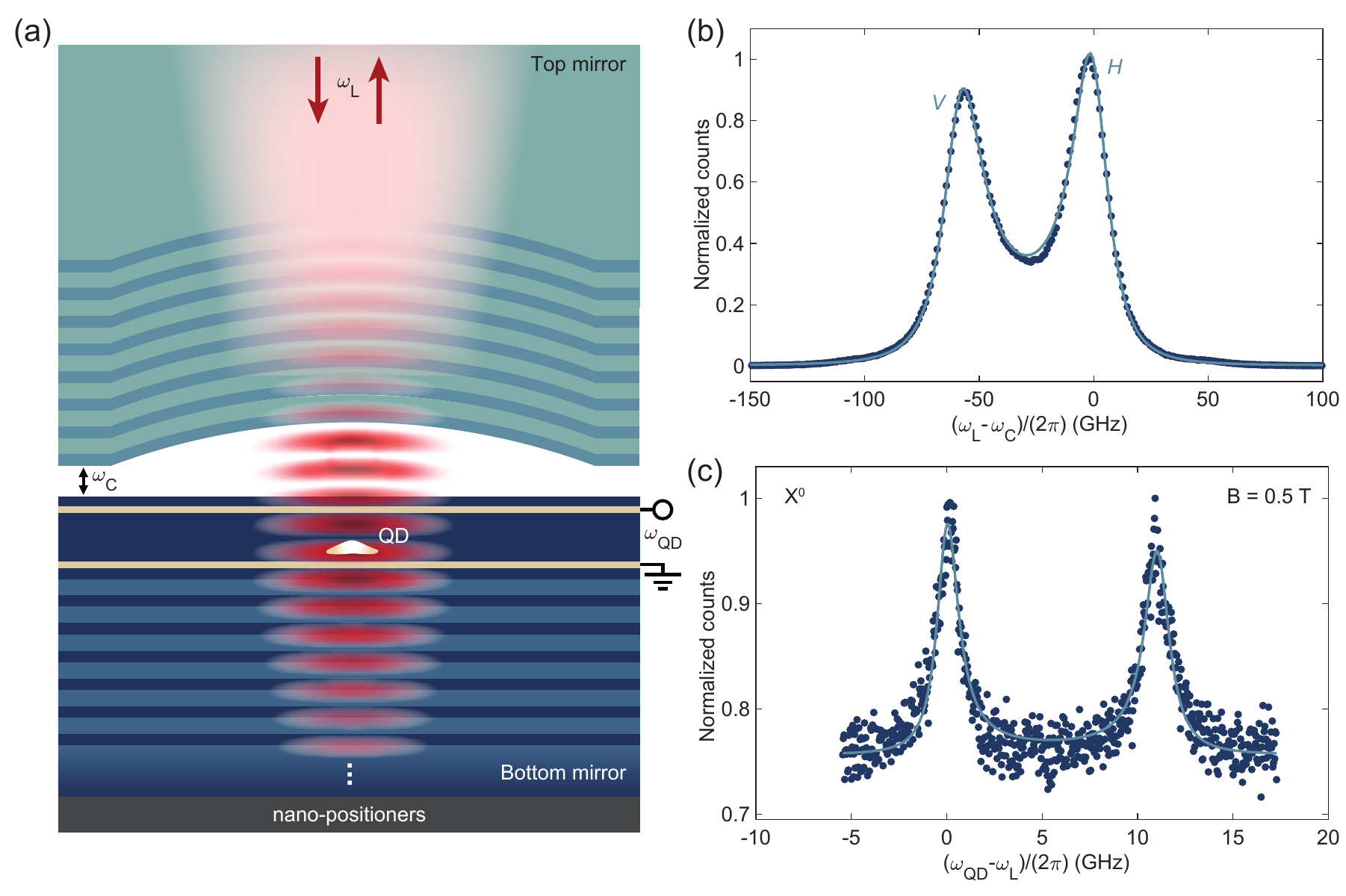}
\caption{\textbf{Microcavity and quantum dot characterisation.} \textbf{(a)} Schematic (adapted from Ref.\,\cite{Tomm2021}) of the QD and one-sided cavity setup. The QDs are embedded in a diode structure which allows tuning $\omega_{\rm QD}$. The cavity is composed of separate top and bottom mirrors, which allow tuning its frequency $\omega_{\rm C}$ with the help of nano-positioners. The laser light pulses with angular frequency $\omega_{\rm L}$ are coupled into the cavity and exit via the top mirror. \textbf{(b)} Normalised signal versus laser detuning of the cavity resonance. The cavity resonance is split into two linearly- and orthogonally-polarised modes, $H$ and $V$, separated by 58.8\,GHz. The $\mathcal{Q}$-factor is measured to be $\mathcal{Q}=16,200\pm1,200$, leading to a cavity loss rate $\kappa/(2\pi)=(20.1\pm1.5)$\,GHz. \textbf{(c)} Normalised signal versus QD detuning from the reference laser ($\lambda=918.23$\,nm). The neutral exciton X$^0$ transitions in the QD are separated by a fine-structure splitting (FSS). Upon application of an out-of-plane magnetic field of B=0.5\,T, the transitions are split further apart, resulting in a $\rm{FSS}=(11.0\pm0.1)$\,GHz. Application of the magnetic field allows us to work with just one of the QD transitions (the lower-frequency one).}
\label{SIfig:cavQDscans}
\end{figure*}

The open-access nature of the cavity allows the tuning of its optical resonance $\omega_{\rm C}$ with the help of nano-positioners. One can also change the frequency of the input laser light $\omega_{\rm L}$ which enters and exits the system. A cavity scan is shown in Supplementary Fig.\,\ref{SIfig:cavQDscans}b, where the evident splitting of the fundamental mode into two orthogonal $H$ and $V$ linearly-polarised modes can be observed~\cite{Tomm2021PRAppl}, $\rm{MS}=(\omega_{\rm V}-\omega_{\rm H})/(2\pi)=-58.8$\,GHz. The mode-splitting is caused by residual strain in the semiconductor heterostructure. By fitting a double-Lorentzian line one extracts the total cavity loss rate $\kappa/(2\pi)=(20.1\pm1.5)$\,GHz, equivalently a quality factor $\mathcal{Q}=16,200\pm1,200$. Upon application of a voltage across the gates of the diode, the QD transition frequencies $\omega_{\rm QD}$ can be tuned using the DC-Stark effect. We use the transitions of a neutral exciton $X^0$, which exhibits a so-called fine-structure splitting (FSS): there are two non-degenerate, linearly-polarised dipole moments. In order to work with one transition only, we search for a highly strained region in the sample in which both the FSS of the QD and the mode-splitting of the cavity are large. We select a QD with a starting linear frequency splitting $\rm{FSS}=(7.6\pm0.1)$\,GHz. We apply an out-of-plane magnetic field B=0.5\,T, large enough to push the transitions further apart via the Zeeman effect, but not large enough to influence significantly the selection rules. (In a large magnetic field, the Zeeman splitting becomes much larger than the FSS such that the transitions become circularly polarised.) With this magnetic field strength, we finally have $\rm{FSS}=(11.0\pm0.1)$\,GHz as shown in Supplementary Fig.\,\ref{SIfig:cavQDscans}c. We focus the rest of the discussion on the lower-frequency transition.

\section{Coupled quantum dot-cavity system}

We characterise the coupled QD-cavity system. We use input-output theory and follow the procedure described in Ref.\,\cite{Auffeves-Garnier2007PRA, Antoniadis2022} to describe the interaction between the QD transition of interest, a two-level system (TLS), and two orthogonally and linearly polarised cavity modes. We examine the case of a continuous-wave, low-power input laser field of amplitude $b_{\rm in}$ with arbitrary polarisation,
\begin{equation}
    b_{\rm in} = \alpha_H b_{\rm in} \mathbf{h} + \alpha_V b_{\rm in} \mathbf{v},
\end{equation}
where $\alpha_{H/V}$ are the complex coefficients of the Jones vectors in the orthonormal basis $\mathbf{h}=(1\;\;0)^\intercal$ and $\mathbf{v}=(0\;\;1)^\intercal$.

We consider a one-sided cavity with two optical modes along $\mathbf{h}$ and $\mathbf{v}$, both coupled to the continuum of modes with decay rate $\kappa_H=\kappa_V=\kappa$. We use the rotating frame of the quantum emitter. The field reflected by the QD-cavity system $b_{\rm r}$ is
\begin{equation}
    b_{\rm r} = \epsilon_H b^H_{\rm r} \mathbf{h} + \epsilon_V b^V_{\rm r} \mathbf{v},
\end{equation}
with the probed reflected field amplitude in each polarisation given by
\begin{equation}
    b^{H/V}_{\rm r} = \alpha_{H/V} \, b_{\rm in} \left(1-t_{H/V}\right) - i \, t_{H/V} \, \Gamma_{H/V} \, \sigma_-,
\end{equation}
where $\sigma_-$ is the expectation value of the Pauli lowering operator for the TLS in the low-power regime. We denote $\Gamma_{H/V}=\frac{4\,g^2_{H/V}}{\kappa}$ as the Purcell-enhanced atomic decay rate due to the interaction, with $g_{H/V}$ the coupling constant between the linear dipole moment of the TLS and each cavity mode. We note that for linear transitions, $g_{H/V}$ is real and given by the projection of the TLS's linear dipole moment onto the cavity mode's linear electric vacuum field. We introduce also the transmission through the cavity modes in the absence of the quantum emitter:
\begin{equation}
    t_{H/V}=\frac{1}{1+2i\left(\frac{\Delta_{H/V}-\Delta_{\rm L}}{\kappa}\right)},
\end{equation}
where $\Delta_{H/V} = \omega^{H/V}_{\rm C}-\omega_{\rm QD}$ is the detuning of the $H/V$ polarised cavity mode with respect to the QD transition, and $\Delta_{\rm L} = \omega_{\rm L}-\omega_{\rm QD}$ is the laser detuning with respect to the QD transition. The lowering operator $\sigma_-$ is responsible for ``coupling'' the two polarised optical modes and described as
\begin{equation}
    \sigma_- = \frac{i \, t_H \, \sqrt{\Gamma_H}\, \alpha_H \, b_{\rm in} +  i \, t_V \, \sqrt{\Gamma_V}\, \alpha_V \, b_{\rm in}}{t_H \, \frac{\Gamma_H}{2} + t_V \, \frac{\Gamma_V}{2} - i \, \Delta_{\rm L} + \frac{\gamma}{2}},
\end{equation}
where we account for the atomic decay rate into non-cavity optical modes via $\gamma$.

Finally, the scattered field amplitude is given by the ratio of reflected field and input field, namely
\begin{equation}
\begin{split}
\label{SIeq:transm}
    t_k = \frac{b_{\rm r}}{b_{\rm in}} = &\, \epsilon_H\,\left[\alpha_H\,\left(1 - 2\,t_H + \frac{2\,t^2_H\,\Gamma_H}{t_H\,\Gamma_H + t_V\,\Gamma_V - 2i \, \Delta_{\rm L} + \gamma}\right)   +   \alpha_V\,\left(\frac{2\,t_H\,\sqrt{\Gamma_H}\,t_V\,\sqrt{\Gamma_V}}{t_H\,\Gamma_H + t_V\,\Gamma_V - 2i \, \Delta_{\rm L} + \gamma}\right)\right]\\
    &+ \epsilon_V\,\left[\alpha_H\,\left(\frac{2\,t_V\,\sqrt{\Gamma_V}\,t_H\,\sqrt{\Gamma_V}}{t_V\,\Gamma_V + t_H\,\Gamma_H - 2i \, \Delta_{\rm L} + \gamma}\right)   +   \alpha_V\,\left(1 - 2\,t_V + \frac{2\,t^2_V\,\Gamma_V}{t_V\,\Gamma_V + t_H\,\Gamma_H - 2i \, \Delta_{\rm L} + \gamma}\right)\right].
\end{split}
\end{equation}

\begin{figure*}[hb!]
\centering
\includegraphics[width=\textwidth]{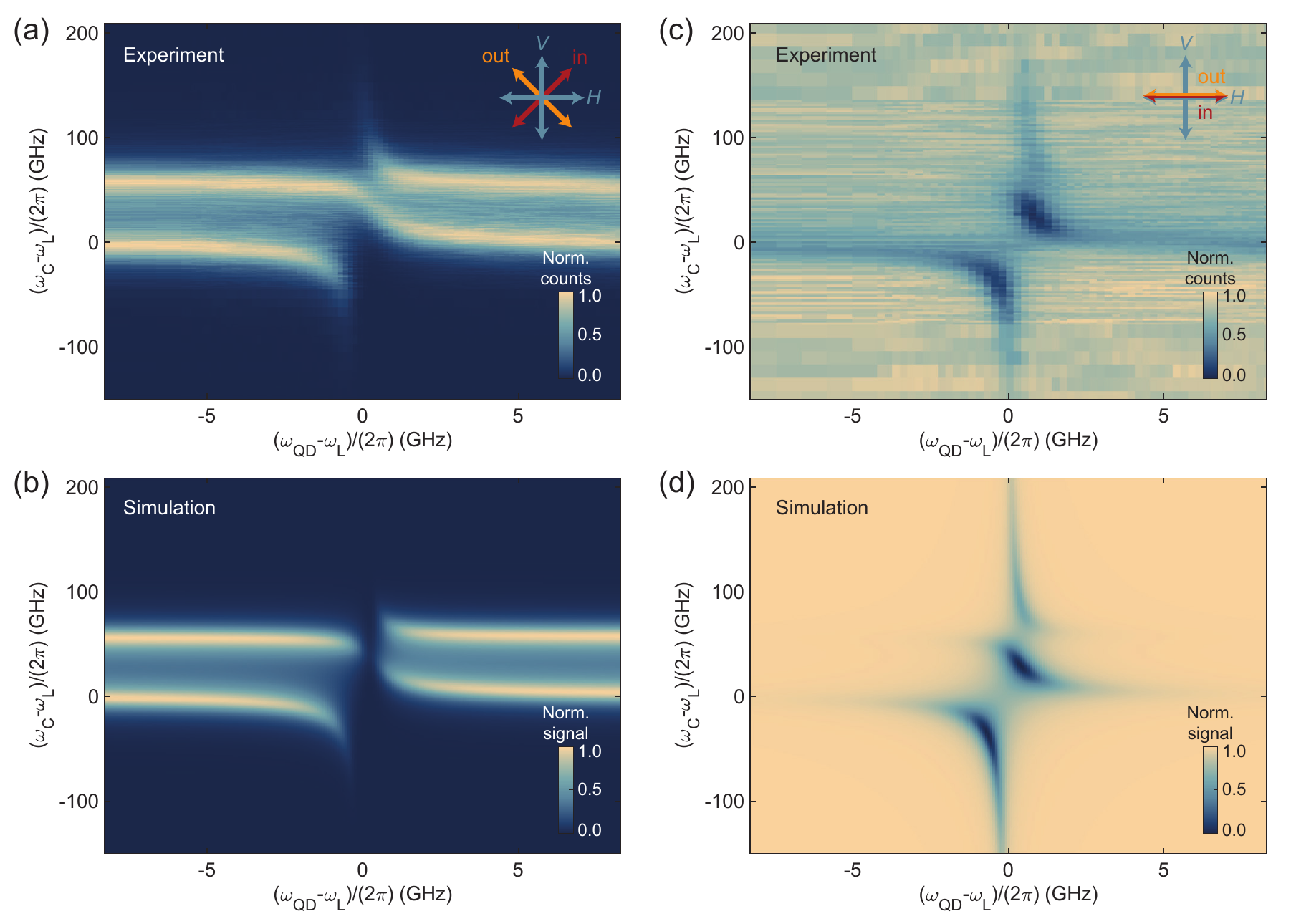}
\caption{\textbf{Resonant response of a QD in a one-side microcavity.} \textbf{(a)} Transmission of laser light through the coupled QD-cavity system as a function of detuning from the cavity and the QD resonances. The measurement was performed in a cross-polarised mode (inset), where the input light is diagonal to the $H$ and $V$ cavity modes and the output polarisation is anti-diagonal. By fitting a Lorentzian function to the cut-through lines at the resonances of the two cavity modes, one can extract the Purcell enhanced decay rate of the QD, $\Gamma_{\rm H}/(2\pi)=4.24$\,GHz and $\Gamma_{\rm V}/(2\pi)=1.30$. \textbf{(b)} Simulation of the transmission through the QD-cavity system using the parameters extracted from the experiment. \textbf{(c)} Spectrum of the back-reflection light from the QD-cavity system. The measurement was performed in co-polarised mode (inset), where input and output modes are horizontally polarised, aligned to the $H$ polarised cavity mode, and \textbf{(d)} respective simulation.}
\label{SIfig:VcVgs}
\end{figure*}

Experimentally, we operate the system with the aid of a cross-polarised microscope\,\cite{KuhlmannRSI2013, Tomm2021}, and we probe the reflected light intensity $\lvert t_k\rvert^2$. We first study the properties of the TLS-cavity interaction in what we call ``transmission'' mode, shown in Supplementary Fig.\,\ref{SIfig:VcVgs}a. Even though the signal is effectively reflected by the system, mathematically it is equivalent to probing the signal transmitted through a two-sided cavity with two orthogonal cavity modes. We achieve this by sending in linear-polarised light ``diagonal" to the cavity modes, and collecting light on the anti-diagonal, as depicted in the inset of Supplementary Fig.\,\ref{SIfig:VcVgs}a. In Eq.\,\ref{SIeq:transm} this means that the input coefficients are $\alpha_H=1/\sqrt{2}$, $\alpha_V=1/\sqrt{2}$, the output coefficients $\epsilon_H=1/\sqrt{2}$, $\epsilon_V=-1/\sqrt{2}$. We extract the Purcell-enhanced decay rate into each mode by taking the full-width-at-half-maximum (FWHM) of the two cavity resonances, $\Gamma_H/(2\pi)=4.24$\,GHz and $\Gamma_V/(2\pi)=1.30$\,GHz. This implies respective coupling constants $g_H/(2\pi)=4.62$\,GHz and $g_V/(2\pi)=2.55$\,GHz. The non-cavity losses are given by the natural decay rate of the emitter, $\gamma/(2\pi)=0.30$\,GHz, a parameter known from previous experiments\,\cite{Tomm2021, Najer2019}. We estimate the Purcell factors $F^H_P=14.1$ and $F^V_P=4.3$. We simulate the response of the system with the parameters determined experimentally and using Eq.\,\ref{SIeq:transm}. The result is shown in Supplementary Fig.\,\ref{SIfig:VcVgs}b. We note that the QD transitions are not perfectly linearly polarised due to Zeeman effect which induces a slight ellipticity to the transitions.

Next, we examine the system in what we call ``back-reflection'' mode, in which both input and output fields are aligned to one of the linearly polarised cavity modes. In this work, we focus on the $H$ cavity; it interacts more strongly with the lower energy transition of the QD. In this case, for the input and output polarisation coefficients, we have $\alpha_H=1$, $\alpha_V=0$ and $\epsilon_H=1$, $\epsilon_V=0$. The experimental and simulated results are presented in Supplementary Fig.\,\ref{SIfig:VcVgs}c,d. We remark that in the case of a TLS with decay rate $\Gamma$ coupled to a single cavity mode with coupling rate $g$ and $\gamma=0$, the amplitude reflection coefficient would be unitary across the entire spectrum. In this limit the $\beta$-factor, defined as $\beta=F_P/(F_P+1)$, is unitary and the amplitude reflection coefficient reduces to Eq.~1 of the main text. At resonance the amplitude reflection coefficient is $-1$, which indicates that the light is scattered with unity intensity but undergoes a $\pi$ phase shift, a useful condition for photon-photon quantum gates\,\cite{Chen2021}. This limit is taken in the theoretical modelling throughout this work, see Sec.\,\ref{SIsec:theory}.

In this work, the unity $\beta$-factor condition is almost achieved. We take advantage of the tunable nature of the miniaturised Fabry-Perot cavity to maximise the $\beta$-factor. Employing xy-nanopositioners one can position the QD relative to the optical axis of the cavity, and in this way tune the $\beta$-factor, and respectively the QD's decay rate/lifetime. In Supplementary Fig.\,\ref{SIfig:GammaLifetime}, we show the extracted $\Gamma/(2\pi)$ (blue dots, left axis) and equivalent emitter lifetime $\tau_{\rm QD}=1/\Gamma$ (red dots, right axis) as a function of the calculated $\beta$ for different lateral positions of the QD relative to the cavity's optical axis. Here, $\beta$ is calculated from the derived coupling constant $g=g_H$ at each position, and the known $\kappa$ and $\gamma$. For all experiments in this manuscript we work in the condition of maximised coupling, with parameters $\beta=0.934$, decay rate $\Gamma/(2\pi)=4.24$\,GHz and respective lifetime $\tau_{\rm QD}=37.5$\,ps. The average photon number per lifetime needed to saturate the quantum emitter in a one-sided cavity\cite{Auffeves-Garnier2007PRA, Antoniadis2022}, i.e.\ the ``critical photon number'' $\bar{n}_c$, is given by $\bar{n}_c=1/(8\beta)=0.1338$.

\begin{figure*}[hb!]
\centering
\includegraphics[width=0.5\textwidth]{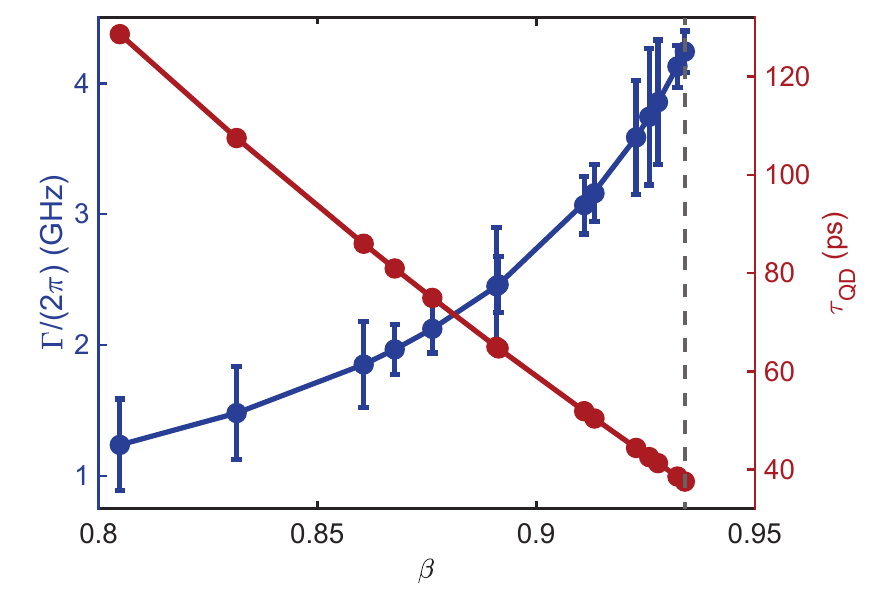}
\caption{\textbf{Tuning the lifetime of the QD coupled to the microcavity.} The open-access microcavity enables tuning the lateral position of the QD relative to the centre of the cavity field, modifying \textit{in-situ} the coupling rate $g$, respectively the $\beta$-factor. We determine experimentally the Purcell-enhanced linewidth of the best-coupled dipole transition $\Gamma=\Gamma_{\rm H}$ (left axis, blue data points) and its respective expected lifetime $\tau_{\rm QD}$ (right axis, red data points). When the QD is well centred with the cavity mode (dashed gray line), $\beta=0.934$, $\Gamma/(2\pi)=4.24$\,GHz and $\tau_{\rm QD}=37.5$\,ps.}
\label{SIfig:GammaLifetime}
\end{figure*}

\section{Experimental setup}
The experimental setup for the generation and detection of single photons and photonic bound states consists of five segments, as shown in Supplementary Fig.\,\ref{SIfig:setup}. We employ a tunable continuous-wave laser (TOPTICA Photonics A.G., Germany) and select the laser power sent to the system with an acousto-optic modulator (AOM) controlled digitally by a PID-loop. We create electric Gaussian-shaped signals with FWHM between 75\,ps and 2\,ns with an arbitrary wave generator (AWG). A trigger signal is sent to a time-correlator to synchronize the experiment, and the analog output from the AWG is used to modulate an electro-optic modulator (EOM). The generated light pulses are sent into a 99:1 fibre beam-splitter, which we employ as an approximate optical circulator. We send 1\% of the light intensity into the QD-cavity system, which resides in a helium bath-cryostat at a temperature of 4.2\,K. The light is back-scattered and 99\% of it is sent to a 50:50 fibre-beam splitter and finally to two superconducting nanowire single-photon detectors (SNSPDs, Single Quantum B.V., The Netherlands). The time-tags of the single-photon detection events are recorded with a correlator (Time Tagger, Swabian Instruments, Germany). In the experimental determination of a three-photon correlation function $G^{(3)}$, we divide the signal once more with a second 50:50 fibre-beam splitter and use four SNSPD detectors.

\begin{figure*}[ht!]
\centering
\includegraphics[width=\textwidth]{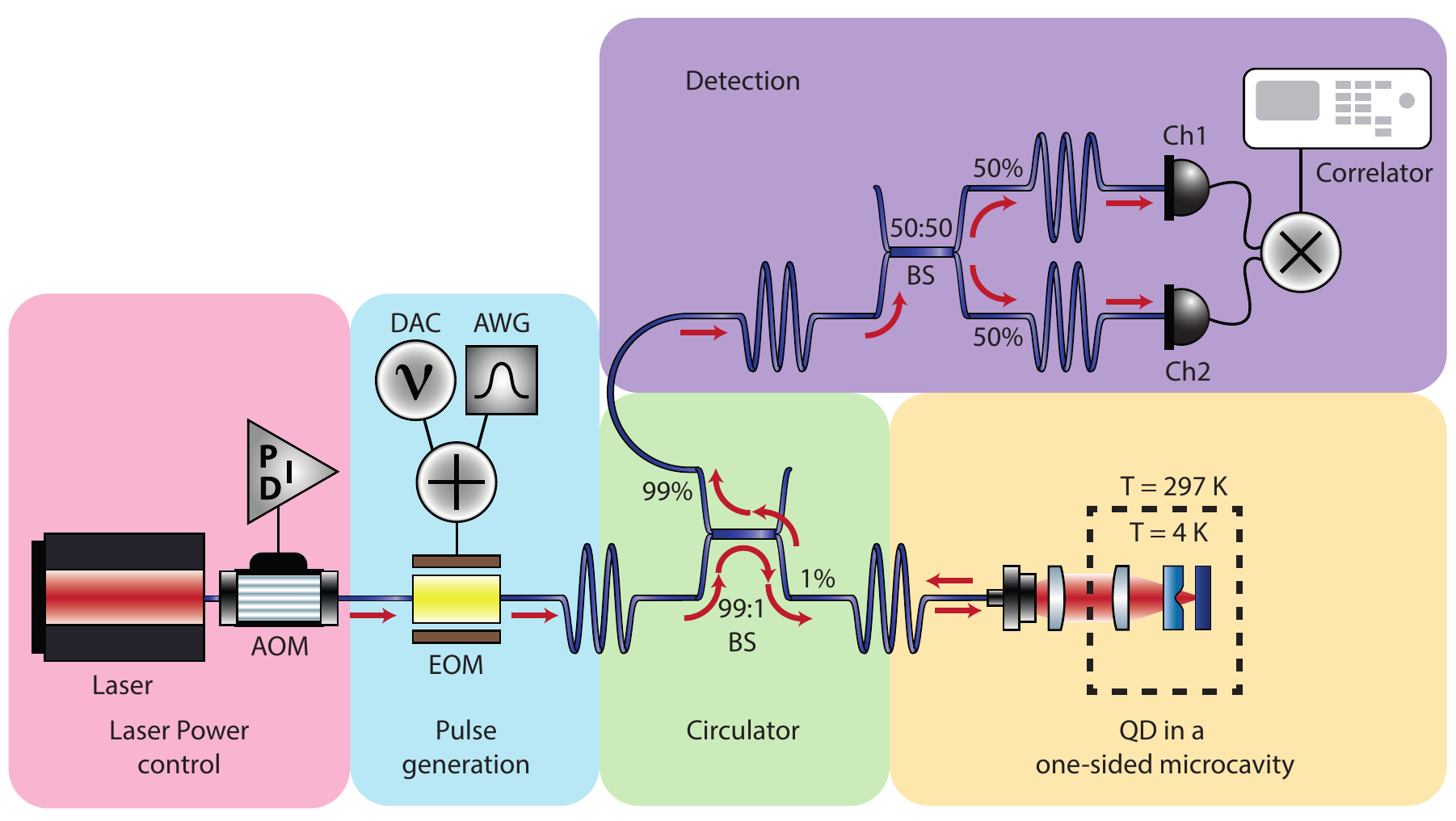}
\caption{\textbf{The experimental setup.} Overview of the experimental setup for the measurement of the single photon and two-photon scattering dynamics. Light from a CW laser passes through an AOM (acousto-optic modulator) controlled via a PID loop in order to control the laser power. Laser light is sent to an EOM (electro-optic modulator) in order to create Gaussian pulses. These pulses are sent to the QD in a one-sided microcavity via a mock-circulator composed of a 99:1 fibre BS (beam-splitter). In the circulator, 1\% of the input light is sent to the QD-cavity system, but 99\% of the back-reflected light (co-polarised) is sent to the detection setup consisting of a Hanbury Brown-Twiss setup (a 50:50 BS with two detectors, Ch1 and Ch2, and a correlator).}
\label{SIfig:setup}
\end{figure*}

\section{Wigner delay induced by a one-sided cavity}
\label{SIsec:cavityWdelay}
The elastic scattering of a wavepacket by a resonator is not an instantaneous process. It was shown by E.P.~Wigner in 1955 that the scattered wave propagates with a time delay with respect to an unscattered wave~\cite{Wigner1955}. The so-called ``Wigner delay" is given by the derivative of the phase shift acquired by the wave with respect to its frequency~\cite{Bourgain2013}. The delay is frequency dependent. In the case of a one-sided optical cavity with an inverse lifetime $\kappa$ the Wigner delay inherited by the interacting pulse is given by
\begin{equation}
\label{SIeq:cavWignerdelay}
    \Delta\tau_C(\omega_L,\omega_C) = \frac{4}{\kappa}\,\frac{1}{1+4\left(\frac{\omega_{\rm L}-\omega_{\rm C}}{\kappa}\right)^2}.
\end{equation}
It is easy to see that at resonance ($\omega_{\rm L}=\omega_{\rm C}$) the Wigner delay reduces to $\Delta\tau_C(0,0) = 4/\kappa$.

\begin{figure*}[h!]
\centering
\includegraphics[width=\textwidth]{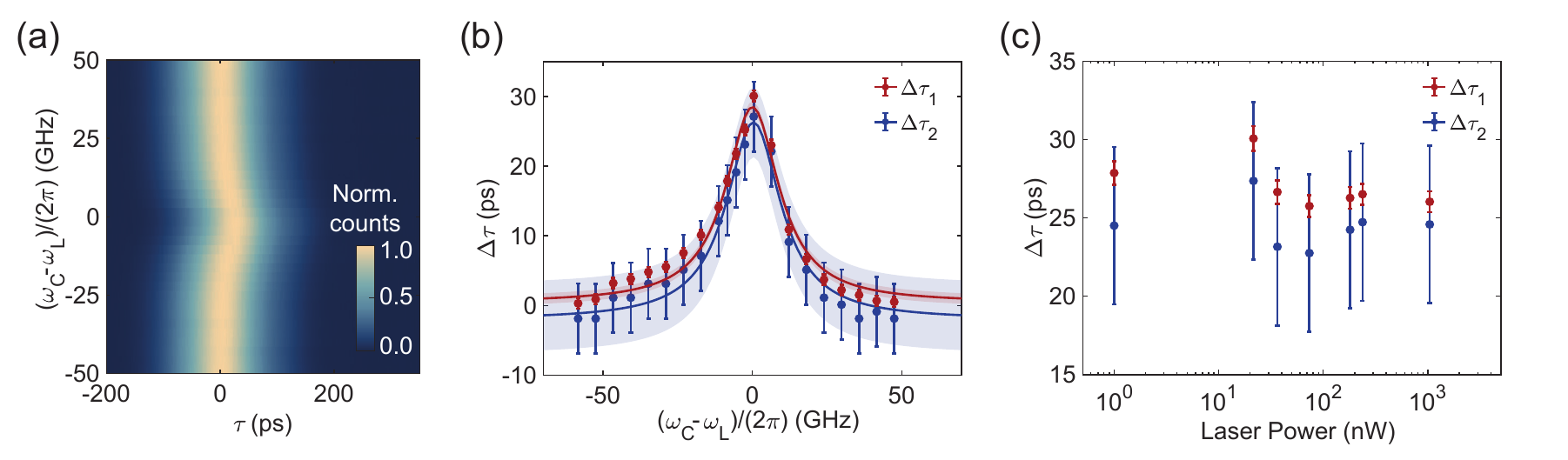}
\caption{\textbf{Wigner delay induced by a one-sided cavity.} \textbf{(a)} Measured $G^{(1)}(\tau)$ histogram as a function of cavity detuning. \textbf{(b)} Measured pulse peak delay $\Delta\tau$ of $G^{(1)}$ (red data points) and $G^{(2)}$ (blue data points) as a function of cavity detuning. For an optical cavity (a classical resonator), the Wigner delay is photon-number independent, and at resonance equal to $\Delta\tau_C=4/\kappa$ for a one-sided cavity. By fitting Eq.\,\ref{SIeq:cavWignerdelay} (solid lines) we determine $\kappa_{\rm fit}/(2\pi)=(21.6\pm0.2)$\,GHZ. \textbf{(c)} The Wigner delay for single-photon states (red data points) and two-photon states (blue data points) at resonance measured for different laser powers: the delay is independent of the power.}
\label{SIfig:cavitydelay}
\end{figure*}

We study the dynamical response of a $\sim$135\,ps Gaussian pulse scattered by the one-sided cavity. We work in ``back-reflection'' mode, such that the input light only interacts with the $H$ cavity mode. Supplementary Fig.\,\ref{SIfig:cavitydelay}a shows the time of arrival of the pulse $G^{(1)}(\tau)$ as a function of detuning between the laser and the cavity resonance. Even for the shortest pulses used here, the scattered pulse is shape-maintaining over the entire spectrum. We evaluate the delay of the pulse peak $\Delta\tau$ for one- (red dots) and two-photon components (blue dots) as a function of detuning (Supplementary Fig.\,\ref{SIfig:cavitydelay}b). The delay experienced in both cases is the same, and described by Eq.\,\ref{SIeq:cavWignerdelay}, which we fit to the data (red and blue solid lines) allowing us to retrieve the cavity linewidth $\kappa_{\rm fit}/(2\pi)=(21.6\pm0.2)$\,GHZ. This value is the same (within the error bars) as the value of $\kappa$ determined via the intensity measurement described in Section\,\ref{SIsec:cavityQDcharacterisation}, $\kappa/(2\pi)=(20.6\pm1.5)$. The photon lifetime in the cavity is $\tau_{\rm C}=1/\kappa_{\rm fit}=(7.3\pm0.1)$\,ps, leading to a Wigner delay at resonance of $\Delta\tau_{\rm C}(0,0) = 4/\kappa_{\rm fit}=(29.2\pm0.4)$\,ps. Finally, we show in Supplementary Fig.\,\ref{SIfig:cavitydelay}c that the Wigner delay for a classical resonator is linear, i.e.\ independent of input laser power. For the experiments in the main manuscript we use the average of the delays extracted in this measurement as the ``classical" (i.e.\ cavity-only) baseline.

\section{Properties of the photon-number dependent scattering}

\begin{figure*}[b!]
\centering
\includegraphics[width=\textwidth]{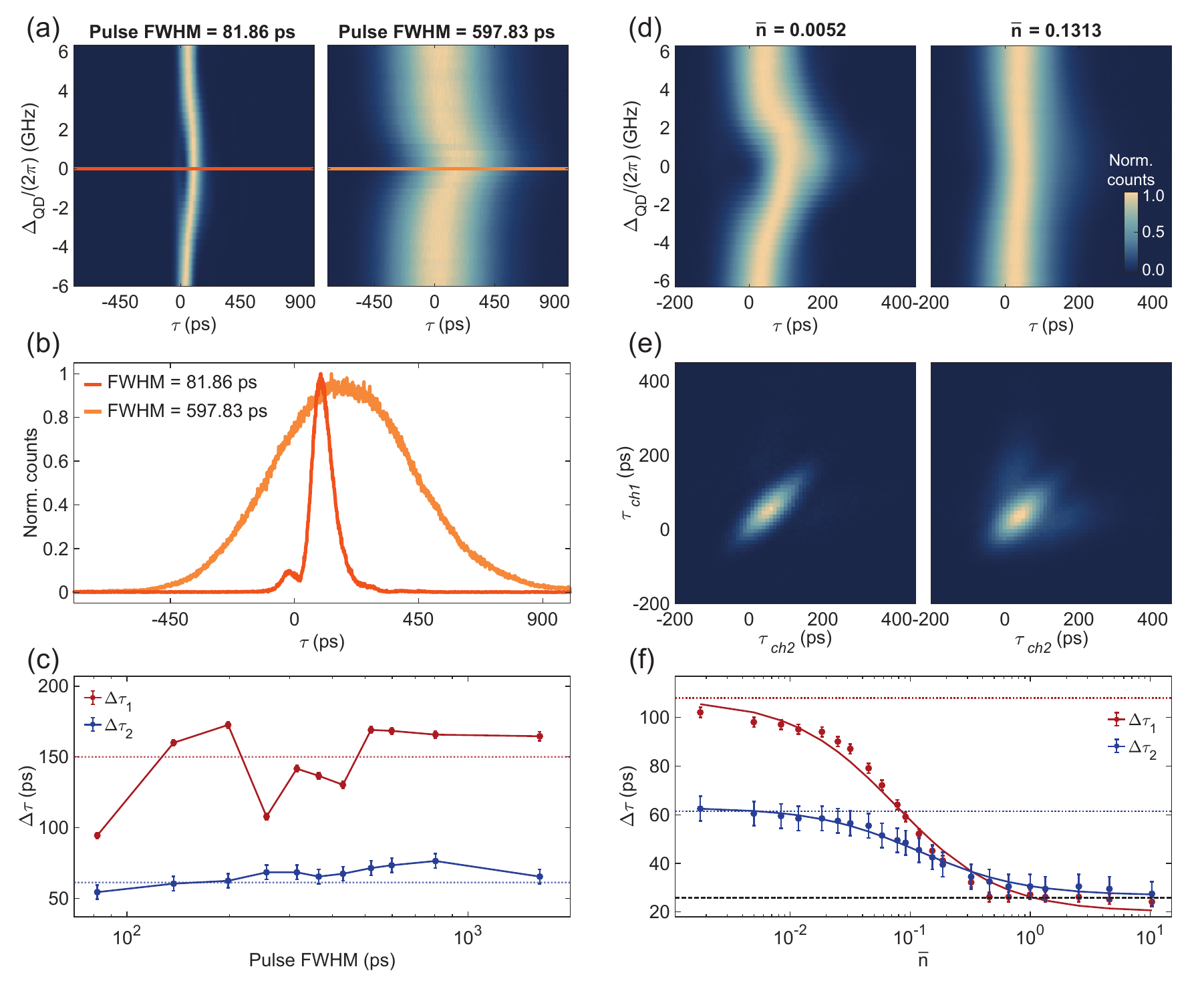}
\caption{Input pulse width and input power dependent dynamics. (a) Time of arrival $\tau$ of the propagated pulse power $G^{(1)}(\tau_{\rm ch1})$ as a function of QD detuning from the laser resonance $\Delta_{\rm QD}/(2\pi)$ for a short (pulse FWHM = 81.86\,ps, left) and a long (pulse FWHM = 597.83\,ps, right) input Gaussian pulse. (b) Line cut-through of (a) at resonance, showing the pulse deformation for short pulses. (c) Peak delay $\Delta\tau$ as a function of input pulse width for one (red) and two photons (blue). The red and blue horizontal dashed lines represent the theoretical values in the infinite-pulse-width limit. (d) Time of arrival $\tau$ of the propagated pulse power $G^{(1)}(\tau_{\rm ch1})$ for a 135\,ps Gaussian pulse as a function of QD detuning from the laser resonance for an average input photon number $\bar{n}=0.0052$ (left) and $\bar{n}=0.1313$ (right), and (e) respective $G^{(2)}(\tau_{\rm ch1},\tau_{\rm ch2})$ maps at resonance. (f) Peak delay as a function of average input photon number for one- (red) and two-photon (blue) states. Solid lines are fits to the TLS-saturation relation, with critical photon number $\bar{n}_c=0.1338$. The red and blue dashed lines show the modelled $\Delta\tau$ for one- and two-photon states for a 135\,ps Gaussian pulse in the low power limit, and the black dashed line is the delay of the cavity alone.}
\label{SIfig:scattproperties}
\end{figure*} 

The scattered light dynamics depend on the input pulse shape and temporal width. Shorter pulses in time, a few times the lifetime of the emitter, undergo higher-order distortion as more spectral components interact with the TLS (see Sec.\,\ref{SIsec:theory}). Much longer pulses are spectrally narrow and the continuous-wave (CW) limit is appropriate in which only a delay (phase shift) is impinged onto the pulse. In Supplementary Fig.\,\ref{SIfig:scattproperties}a, the normalised power $G^{(1)}(\tau)$ as a function of QD--laser detuning ($\Delta_{\rm QD}=\omega_{\rm QD}-\omega_{\rm L}$) is shown. The laser is in resonance with the cavity. We show the one-photon scattering response for two pulse widths, a shorter one (left, pulse intensity-FWHM of 81.86\,ps, corresponding to $\sigma\Gamma=1.3$) and a longer one (right, pulse intensity-FWHM = 597.83\,ps, corresponding to $\sigma\Gamma=9.6$). In both cases, far from resonance, the pulse is delayed by the classical delay of the cavity alone and does not present any reshaping. In the shorter-pulse limit, the transmitted pulse is distorted near the resonance and no longer corresponds to a Gaussian pulse, as is seen in Supplementary Fig.\,\ref{SIfig:scattproperties}b. The distortion of the pulse results in a deviation of the peak delay $\Delta\tau$ in shorter pulses compared to the infinite-pulse limit, in which the distortion is negligible. The two-photon auto-correlation maps for these pulse widths are discussed in the main text. We probe the peak delay at resonance for single photons $\Delta\tau_1$ and for two-photon states $\Delta\tau_2$ as a function of pulse FWHM, displayed in Supplementary Fig.\,\ref{SIfig:scattproperties}c, where the red and blue dashed lines correspond to the theoretical prediction neglecting the effects of distortion. We point out that the single-photon scattering dynamics are more sensitive to dispersion effects (see Fig.\,2 of main text) than the dynamics of the two-photon bound states; this explains the larger variance in $\Delta\tau_1$ in this analysis.

The response of a single TLS is highly susceptible to the number of photons impinging on it, as the TLS saturates upon absorption of a single photon\,\cite{Chang2007,Auffeves-Garnier2007PRA}. This results in a strong nonlinear response even at very low input laser powers. We expect the few-photon nonlinearity to fade with strong coherent fields, as in this regime a large fraction of the pulse is scattered without interacting with the saturated TLS. Supplementary Fig.\,\ref{SIfig:scattproperties}d presents the normalized power-time response as a function of QD-detuning from the laser-cavity resonance for a pulse with $\sim$135\,ps FWHM in the case of a low- (2.89\,nW, left) and high-power (73.64\,nW, right) input. These laser powers (measured at the ``power control'' setup) translate to an average photon-number per lifetime of $\bar{n}=0.0052$ and $\bar{n}=0.1313$ at the cavity, respectively. The QD saturates at the critical photon number \cite{Antoniadis2022} $\bar{n}_c=1/(8\beta)=0.1338$. In the latter measurement, the reduced delay at resonance testifies to the saturation of the QD. The fingerprints of two-photon correlated states (clustering and delay of coincidence counts along the diagonal) also diminish (Supplementary Fig.\,\ref{SIfig:scattproperties}e). We present in Supplementary Fig.\,\ref{SIfig:scattproperties}f the peak delay observed for one (red) and two photons (blue) as a function of average photon number per lifetime. At low photon-number, the measurements correspond well to the simulated $\Delta\tau$ for this pulse width (red and blue dashed lines), converging gradually as the laser power increases to the limit where only the classical delay, induced by the cavity alone, remains (black dashed line) as the laser power increases. The delay response for both one- and two-photon states correspond well to a TLS-saturation power law like function,
\begin{equation}
    \Delta\tau(\bar{n})= 1 - (1-\Delta\tau^\infty/\Delta\tau^0)\cdot\frac{\bar{n}/\bar{n}_c}{1+\bar{n}/\bar{n}_c},
\end{equation}
where $\Delta\tau^0$ and $\Delta\tau^\infty$ are the peak delays measured in the limits of zero and infinite input photons, respectively. The fits are depicted with red and blue solid lines in the plot.

Next, we evaluate $G^{(1)}(\tau)$ and the diagonals of the auto-correlation functions $G^{(2)}(\tau,\tau)$ and $G^{(3)}(\tau,\tau,\tau)$ and determine the dynamical properties of the photon-number dependent scattering. We present in Supplementary Fig.\,\ref{SIfig:propertiesvsn}a,b the values of peak delay $\Delta\tau$ and relative pulse width $\sigma_n/\sigma_{\ket{in}}$ as a function of photon-number $n$. For each $n$, these values are extracted from $G^{(1)}(\tau)$, $G^{(2)}(\tau,\tau)$ and $G^{(3)}(\tau,\tau,\tau)$. Here, we extract the average and variance from all experiments presented in the course of this work under low-power resonant conditions.

\begin{figure*}[h!]
\centering
\includegraphics[width=\textwidth]{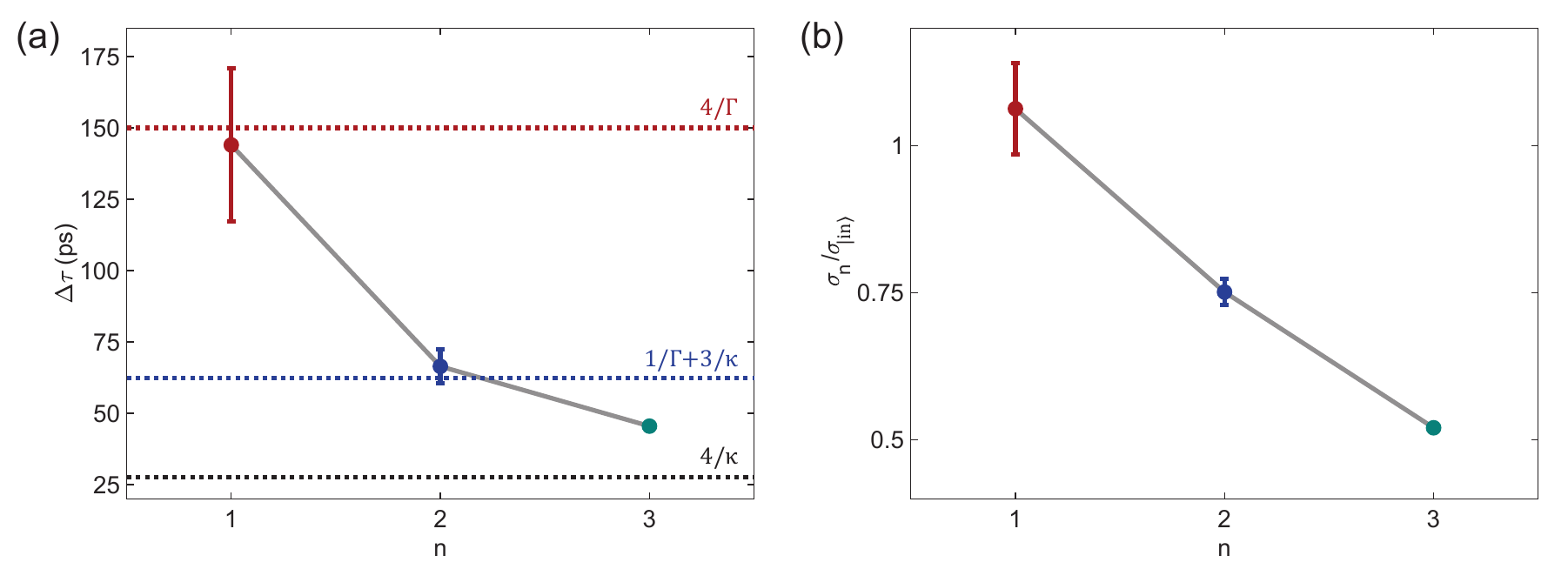}
\caption{\textbf{Photon-number-dependent scattering dynamics.} \textbf{(a)} Average peak delay of the scattered pulse $\Delta\tau$, and \textbf{(b)} average ratio of the measured pulse width after scattering ($\sigma_n$) to the input pulse width $\sigma$, for different photon-numbers $n$. The values are extracted by averaging over all the low-power resonant experiments carried out in the course of this work (including differing pulse widths), probed via $G^{(1)}$, $G^{(2)}(\tau,\tau)$ and $G^{(3)}(\tau,\tau,\tau)$.}
\label{SIfig:propertiesvsn}
\end{figure*}

The single-photon components undergo a delay $\Delta\tau_1 = (144.02\pm27.90)$\,ps, in good agreement with the predicted value of $\Delta\tau^{\rm theory}_1 = 4/\Gamma = 150.00$\,ps. The large error stems from the sensitivity of the peak delay to distortions, as discussed in the main text and Section\,\ref{SIsec:theory}; the distortions depend sensitively on the exact detunings which can drift slightly. We determine the two- and three-photon delays to be $\Delta\tau_2 = (66.48\pm5.97)$\,ps and $\Delta\tau_3 = (45.51\pm0.09)$\,ps. The two-photon delay corresponds well to the prediction of $\Delta\tau^{\rm theory}_2=1/\Gamma+3/\kappa=61.3$\,ps. We note that as we approach the classical limit of $n\rightarrow\infty$, the delay converges to the value induced by the classical cavity alone (see Section\,\ref{SIsec:cavityWdelay}). Finally, we compare the Gaussian width of the interacting $n$-photon output pulse and the input pulse and determine $\sigma_1/\sigma_{\ket{in}} = 1.06\pm0.08$, $\sigma_2/\sigma_{\ket{in}} = 0.75\pm0.02$ and $\sigma_3/\sigma_{\ket{in}} = 0.52\pm0.00$.

\section{Theory}
\label{SIsec:theory}

In the Supplement here, we provide a detailed derivation of the equations used to compute the results in the main text. Here we provide a scheme to compute the output state when considering a weak coherent input pulse with average photon number $\bar n \ll 1$. We use one- and two-photon scattering matrices that have previously been computed \cite{Shi2011PRA, Rephaeli2012IEEE}. In addition to this we compute the one- and two-photon delay and dispersion a pulse undergoes when interacting with a single-sided cavity QED system. In particular we do this for the one-photon pulse and we do this in the two-excitation center-of-mass coordinate for the two-photon bound state, whose form we also compute.

The TLS-cavity system can be described by the Jaynes-Cummings interaction, where the cavity is unidirectionally coupled to a one-dimensional continuum of modes, specified by the Hamiltonian
\begin{equation}
     \hat{\mathcal{H}}= -i\,v_g\,\int dx\,\hat{a}^\dagger(x)\,\partial_x \hat{a}(x) + \,\Delta_C\,\hat{c}^\dagger\hat{c}  +  \eta\,\left(\hat{a}(0)\hat{c}^\dagger + \hat{c}\hat{a}^\dagger(0)\right) + g\,\left(\hat{c}\hat{\sigma}_+ + \hat{\sigma}_-\hat{c}^\dagger\right).
\end{equation}
The first two terms describe respectively the free-evolution of the waveguide mode at position $x$ with annihilation operator $\hat{a}$ and the cavity field with annihilation operator $\hat{c}$ and cavity-atom detuning $\Delta_C = \omega_C - \omega_0$. Here, the energy has been renormalised to that of the atom and the dispersion of the mode continuum has been linearised about this point, and $v_g$ is its group velocity. The last two terms express the waveguide-cavity exchange with coupling constant $\eta = \sqrt{\kappa v_g}$, where $\kappa$ is the cavity decay rate, and the cavity-TLS interaction with coupling rate $g$. Here $\hat{\sigma}_-$ ($\hat{\sigma}_+$) is the Pauli lowering (raising) operator

\subsection{The input state}

We truncate the input coherent pulse with real-valued amplitude $\alpha = \sqrt{\bar n}$ at two photons,
\begin{equation}
| \textrm{in} \rangle \sim e^{- \alpha^2/2}\left[ 1  + \alpha \hat{a}_{\rm in}^\dagger + \frac{\alpha^2}{2} \hat{a}_{\rm in}^\dagger \hat{a}_{\rm in}^\dagger \right] | 0 \rangle.
\end{equation}
Here, 
\begin{equation}
\hat{a}^\dagger_{\rm in} = \int dx \hat{a}^\dagger(x) f(x)
\end{equation}
creates a photon in the input mode with normalized spatial profile $f(x)$ (with equivalent annihilation operators). The input mode profile produced by the laser is approximately Gaussian. For an ideal Gaussian pulse with width $\sigma$ and central frequency detuning from the atomic resonance $v_g k_0$ we have
\begin{equation}
\label{eq:inputPulseShape}
f(x) = \frac{1}{\sqrt{\sigma}\pi^{1/4}} e^{i k_0 x} e^{- \frac{x^2}{2 \sigma^2}}.
\end{equation}
Here, $v_g$ is the group velocity of the mode that is coupled to the cavity (see Fig. 1 of the main text for a schematic of the setup). In our experimental setup we have a free space mode with $v_g=c$.

\subsection{Scattering Matrix}

To compute the output state we apply photon scattering matrices on the one- and two-photon components of the coherent state. The one- and two-photon scattering matrices for a single-sided cavity QED setup have previously been reported \cite{Shi2011PRA, Rephaeli2012IEEE}. The single-photon scattering matrix can be written as a transmission coefficient in $k$-space such that scattering a photon maps
\begin{equation}
\label{eq:onePhotScattMat}
\hat{a}^\dagger(k) \rightarrow t_k \hat{a}^\dagger(k) = \frac{v_g k - \Delta_{\rm C} - g^2/(v_g k) - i \kappa/2}{v_g k - \Delta_{\rm C} - g^2/(v_g k) + i \kappa/2} \hat{a}^\dagger(k),
\end{equation}
where $\hat{a}(k)$ is the Fourier Transform of $\hat{a}(x)$. We note that because there is no loss in the system $|t_k|=1$.

The two-photon scattering matrix is given by \cite{Shi2011PRA, Rephaeli2012IEEE} and maps
\begin{equation}
\begin{split}
\label{eq:twoPhotScattMat}
&\int dp_1 dp_2 \hat{a}^\dagger (p_1)\hat{a}^\dagger (p_2) | 0 \rangle f(p_1) f(p_2) \rightarrow  \int dk_1 dk_2 \hat{a}^\dagger (k_1)\hat{a}^\dagger (k_2) | 0 \rangle t_{k_1}t_{k_2}f(k_1) f(k_2) + \\
&\frac{i \sqrt{\kappa} g}{\pi} \int dk_1 dk_2 \hat{a}^\dagger (k_1)\hat{a}^\dagger (k_2) | 0 \rangle s_{k_1}^a s_{k_2}^a \int dp_1 dp_2 f(p_1) f(p_2) \delta(E_p - E_k) \frac{2g (s_{p_1}^c + s_{p_2}^c) + (E_p - 2\omega_c + i \kappa)(s_{p_1}^a + s_{p_2}^a)}{(E_p - \lambda_{+})(E_p - \lambda_{-})},
\end{split}
\end{equation}
where $E_p = p_1 + p_2$ and $E_k = k_1+k_2$ denote the total input and output wavenumbers which must be equal due to conservation of energy, and
\begin{equation}
\begin{split}
s_{k}^a &= \frac{\sqrt{\kappa} \, g}{(v_g k - \Delta_{\rm C} + i \kappa/2)v_g k - g^2},\\
s_{k}^c &= \frac{\sqrt{\kappa} \, v_g k}{(v_g k - \Delta_{\rm C} + i \kappa/2)v_g k - g^2}.
\end{split}
\end{equation}
In addition, we have
\begin{equation}
\lambda_\pm = \frac{2(\omega_0+\omega_c) + \Delta_{\rm C} - 3 i \kappa/2}{2} \pm \sqrt{\left(\frac{\Delta_{\rm C} - i \kappa/2}{2} \right)^2 + 2g^2 }.
\end{equation}
We use (\ref{eq:twoPhotScattMat}) to compute the full two-photon output wavepacket in the main text. The unnormalized second-order correlation function is, in the low-input power limit, proportional to the two-photon wavefunction
\begin{equation}
G^{(2)}(x_1, x_2) = \bar{n}^2 |\psi_2(x_1, x_2)|^2 + O(\bar{n}^3).
\end{equation}
Here, $\psi_2(x_1, x_2)$ is the two-photon wavefunction expressed in real space and $O(\bar n^3)$ indicates terms of order $\bar n^3$ and higher.

\subsubsection{Single-photon scattering, delay, and dispersive effects} 

With the single-photon transmission coefficient one can calculate the delay and dispersion the cavity QED system imparts on an incoming Gaussian single-photon pulse. For an input state
\begin{equation}
|\textrm{in} \rangle = \frac{1}{\sqrt{\sigma}\pi^{1/4}} \int dx \hat{a}^\dagger(x) | 0 \rangle  e^{i k_0 x} e^{- \frac{x^2}{2 \sigma^2}},
\end{equation}
we calculate the output state by taking its Fourier Transform, computing the single-photon scattering using (\ref{eq:onePhotScattMat}), and then inverse Fourier Transforming back to real-space. From this we get the output state
\begin{equation}
| \textrm{out} \rangle = \sqrt{\frac{\sigma}{2\pi}} \frac{1}{\pi^{1/4}} \int dx \hat{a}^\dagger(x) | 0 \rangle \int dk \, t_k  e^{-(k-k_0)^2 \sigma^2/2}e^{i k x}.
\end{equation}
The full output state is computed by calculating the Fourier integral over $k$ numerically.

We get analytic insight into the delay and dispersive properties of pulse propagation by Taylor expanding the transmission coefficient about the detuning $\Delta_{\rm L} = v_g k_0$ at which the Gaussian pulse is centered. Noting that $|t_k|=1$, we write $t_k = e^{i \phi(\Delta_{\rm L}))}$, which implies
\begin{equation}
\phi(\Delta_{\rm L}) = -i \log{(t_{k})}.
\end{equation} 
We then consider a Taylor expansion of $\phi(\Delta_{\rm L})$. Keeping terms up to third order in the Taylor expansion we obtain for the output state
\begin{align}
\label{eq:singleOut}
|\textrm{out} \rangle &\sim \sqrt{\frac{\sigma}{2\pi}} \frac{1}{\pi^{1/4}} \int dx \hat{a}^\dagger(x) | 0, 0, g \rangle \int dk \, e^{i \left[\phi(\Delta_{\rm L}) + (k-k_0)\phi'(\Delta_{\rm L}) + \frac{(k-k_0)^2}{2}\phi''(\Delta_{\rm L}) + \frac{(k-k_0)^3}{3!}\phi'''(\Delta_{\rm L}) \right]}  e^{-(k-k_0)^2 \sigma^2/2}e^{i k x}\\
&= \sqrt{\frac{\sigma}{2\pi}} \frac{1}{\pi^{1/4}} \int dx \hat{a}^\dagger(x) | 0, 0, g \rangle e^{i (\phi(\Delta_{\rm L}) - k_0\phi'(\Delta_{\rm L}))} \int dk e^{i k [x+\phi'(\Delta_{\rm L})]} e^{- \frac 1 2 (k-k_0)^2 \left( \sigma^2 + i \phi''(x) \right) } e^{i \frac{(k-k_0)^3}{3!}\phi'''(\Delta_{\rm L})}.
\end{align}
The different derivatives therefore affect the Gaussian pulse in different ways. The first derivative $\phi'(\Delta_{\rm L})$ describes the delay the pulse undergoes, while the second derivative $\phi''(\Delta_{\rm L})$ describes the broadening and chirp of the Gaussian pulse. These two terms preserve the Gaussian shape of the pulse. On the other hand the third derivative $\phi'''(x)$ describes pulse distortion, whereby the pulse becomes non-Gaussian in shape. These derivatives can be computed analytically using the definition of $t_k$ in (\ref{eq:onePhotScattMat}). The first and second derivatives as a function of detuning $\Delta_{\rm L}$ are,
\begin{align}
\label{eq:singlePhotonDelay}
\phi'(\Delta_{\rm L}) &= \frac{4 \kappa (4 \Delta_{\rm L}^2 + \Gamma \kappa)}{16 \Delta_{\rm L}^4 + \Gamma^2 \kappa^2 + 4 \Delta_{\rm L}^2 \kappa(\kappa - 2\Gamma)},\\
\phi''(\Delta_{\rm L}) &= -\frac{32 \Delta_{\rm L} \kappa [16 \Delta_{\rm L}^4 + 8 \Gamma \Delta_{\rm L}^2 \kappa + \Gamma \kappa^2 (\kappa-3 \Gamma)]}{(16 \Delta_{\rm L}^4 + \Gamma^2 \kappa^2 + 4 \Delta_{\rm L}^2 \kappa (\kappa -2 \Gamma ))^2)}.
\end{align}
Here we have defined $\Gamma=4g^2/\kappa$. On resonance, when $\Delta_{\rm L}=0$, the derivatives simplify to
\begin{align}
\phi'(0) &= \frac 4 \Gamma,\\
\phi''(0) &= 0,\\
\phi'''(0) &= -\frac{32}{\Gamma^3} + \frac{96 \Gamma}{\Gamma^3 \kappa},
\end{align}
where even-order derivatives are zero due to symmetry. These give significant insight into the role of the cavity. They show that on resonance, given that $\Gamma$ is fixed, the decay rate of the cavity does not influence the delay imparted on the pulse because $\phi'(0)$ is independent of $\kappa$. The delay here is the same as scattering off a bare two-level system \cite{Mahmoodian2020}. On the other hand, the cavity can significantly reduce the pulse distortion term on resonance $\phi'''(0)$. In particular when $\kappa=3 \Gamma$ the pulse distortion term vanishes. This occurs near the onset of strong coupling where the fluorescence spectrum develops a flat top.

In this section we have taken advantage of the fact that, due to conservation of energy, the plane-wave states created by $\hat{a}^\dagger(k)$ are scattering eigenstates with eigenvalue $t_k$. We can thus use this to project an input state on a complete basis of these scattering eigenstates and compute the delay and dispersion imparted by the scattering process. Unfortunately, with the form of the two-photon scattering matrix in (\ref{eq:twoPhotScattMat}) we do not have the eigenstates an we therefore cannot do this for the two-excitation states. In order to perform an equivalent calculation for the two-photon input states we find the two-photon scattering eigenstate.

\subsection{Two-photon Scattering Eigenstates}

We now consider the two-excitation eigenstates. These are significantly more complicated than the plane-wave single-excitation eigenstates. This is because the energy of each individual photon is not necessarily conserved. We follow a process similar to that presented in \cite{Iversen2021PRL, Shen2007PRA}. The technique is reminiscent of computing two-excitation Bethe states in spin chains \cite{Giamarchi2003book}.

We begin by writing an Ansatz that spans the two-excitation sector,
\begin{equation}
\begin{split}
|\psi_2 \rangle = \left\{ \frac{1}{\sqrt{2}} \int dx_1 dx_2 \alpha(x_1, x_2) \hat{a}^\dagger(x_1) \hat{a}^\dagger(x_x) + \int dx_1 \hat{a}^\dagger(x_1) \hat{\sigma}_+ b(x_1) + e \hat{\sigma}_+ \hat{c}^\dagger + \int dx_1 \hat{a}^\dagger (x_1) \hat{c}^\dagger c(x_1) +  c_0 \frac{c^\dagger c^\dagger}{\sqrt{2}} \right\}|0,0,g \rangle,
\end{split}
\end{equation}
where $|0,0,g \rangle$ is the state corresponding to no photons in the waveguide or cavity and the atom in the ground state. Using the Hamiltonian in the main text, Schr\"{o}dinger's equation produces five coupled equations
\begin{align}
\label{eq:alphaEq}
-i v_g \left(\frac{\partial}{\partial x_1} + \frac{\partial}{\partial x_1} \right)\alpha(x_1,x_2) + \frac{\eta}{\sqrt{2}} \left[ c(x_1) \delta(x_2) + c(x_2) \delta(x_1) \right] &= E \, \alpha(x_1,x_2),\\
\label{eq:c0Eq}
2 \Delta_{\rm C} c_0 + \sqrt{2} \eta c(0) + \sqrt{2} g e &=  E c_0,\\
\label{eq:bEq}
- i v_g \frac{d}{d x} b(x) + \eta e \delta(x) + g c(x) & = E b(x),\\
\label{eq:eEq}
\Delta_{\rm C} e + \eta b(0) + \sqrt{2} g c_0 &= E e,\\
\label{eq:cEq}
- i v_g \frac{d}{dx} c(x) + \Delta_{\rm C} c(x) + \frac{\eta}{\sqrt{2}} [\alpha(0,x) + \alpha(x,0) ] + \sqrt{2} \eta \delta(x) c_0 + g b(x) &= E c(x).
\end{align}
Our aim is to solve for the scattering eigenstates, which means we want $\alpha(x_1, x_2)$ in the limits $x_1,x_2 \rightarrow -\infty$ and $x_1,x_2 \rightarrow \infty$, giving the in and out states respectively. We note that due to bosonic symmetry $\alpha(x_1, x_2) = \alpha(x_2, x_1)$.

We follow the standard approach for a two-body problem where we use center-of-mass $x_c = (x_1 + x_2)/2$ and difference $x=x_1-x_2$ coordinates. Using these coordinates, when $x_1$ and $x_2$ are away from the origin, Eq.~{\ref{eq:alphaEq}} simply becomes 
\begin{equation}
- i v_g \frac{\partial}{\partial x_c} \alpha(x_c, x) = E\alpha(x_c, x).
\end{equation}
This has the solution
\begin{equation}
\alpha(x_c, x) = e^{i\frac{E}{v_g} x_c} f(x).
\end{equation}
The center-of-mass evolution follows the form of a plane wave which results from the two-photon energy being a conserved quantity. The physics of the two-photon scattering process therefore lies in the function $f(x)$.

The function $f(x)$ is  piecewise continuous, but has discontinuities in its derivative at $x_1=0$ and at $x_2=0$. Due to bosonic symmetry $f(-x) = f(x)$, we therefore, without loss of generality restrict ourselves to $x_1 > x_2$, which implies $x>0$. Values for $x<0$ can be obtained via symmetry. Using this information we write an Ansatz
\begin{equation}
  f(x) =
  \begin{cases}
                                   f_{\rm in}(x) & 0>x_1>x_2 \\
                                   f_{0}(x) & x_1>0>x_2 \\
  f_{\rm out}(x) & x_1>x_2>0.
  \end{cases}
\end{equation}
With this Ansatz we aim to obtain an equation for $f_0(x)$ in terms of $f_{\rm in} (x)$ and then an equation for $f_{\rm out} (x)$ in terms of $f_0(x)$. To get the former we first integrate Eq.~\ref{eq:alphaEq} from $x_1=-\epsilon$ to $x_1=\epsilon$ giving,
\begin{align}
-i v_g \left(\alpha(0^+,x_2) - \alpha(0^-,x_2) \right) + \frac{\eta}{\sqrt{2}} c(x_2) &=0\\
\label{eq:c_for_f0_fin}
c(x) &= \frac{i v_g \sqrt{2}}{\eta} \left[f_0(-x) - f_{\rm in}(-x) \right] e^{i E x/(2 v_g)}.
\end{align}
We then substitute Eq.~\ref{eq:cEq} in Eq.~\ref{eq:bEq} to eliminate $b(x)$, while noting that in these equations we are in the domain $x <0$. We also make use of the fact that $\alpha(x_1,x_2)$ has bosonic symmetry and is piecewise continuous, i.e., 
\begin{equation}
\alpha(x,0)=\alpha(0,x) = \left[\alpha(0^-,x) + \alpha(0^+,x) \right]/2 = \left[f_{\rm in}(-x) + f_0(-x) \right]e^{i E x/(2 v_g)},
\end{equation}
allowing us to eliminate $c(x)$. After some manipulation, we obtain the differential equation
\begin{equation}
\label{eq:DE1}
f_0''(x) - a f_0'(x) + b f_0(x) = f_{\rm in}''(x) - \tilde{a} f_{\rm in}'(x) + \tilde b f_{\rm in}(x),
\end{equation}
where, 
\begin{align}
a &= -\frac{i}{v_g}\left[E - \Delta_{\rm C} + \frac{i \kappa}{2} \right],\\
b &= \frac{E}{2 v_g^2} \left[ \Delta_{\rm C} - \frac{E}{2} + \frac{2g^2}{E} - i \frac{\kappa}{2} \right],\\
\tilde a &= -\frac{i}{v_g}\left[E - \Delta_{\rm C} - \frac{i \kappa}{2} \right],\\
\tilde b &= \frac{E}{2 v_g^2} \left[ \Delta_{\rm C} - \frac{E}{2} + \frac{2g^2}{E} + i \frac{\kappa}{2} \right].
\end{align}
We have again defined the cavity relaxation rate as $\kappa = \eta^2/v_g$.

We now follow the same procedure to obtain an equation for $f_{\rm out}(x)$ in terms of $f_0(x)$. We first integrate Eq.~\ref{eq:alphaEq} from $x_1 =- \epsilon$ to $x_1 =\epsilon$. This gives
\begin{equation}
\label{eq:c_for_fout_f0}
c(x) = \frac{i \sqrt{2} v_g}{\eta} \left[f_{\rm out} (x) - f_0 (x) \right] e^{i E x/(2v_g)}
\end{equation}
We then substitute Eq.~\ref{eq:cEq} in Eq.~\ref{eq:bEq} to eliminate $b(x)$, but now we work in the domain $x > 0$. This time we make use of 
\begin{equation}
\alpha(x,0)=\alpha(0,x) = \left[\alpha(x,0^-) + \alpha(x,0^+) \right]/2 = \left[f_{\rm 0}(x) + f_{\rm out}(x) \right]e^{i E x/(2 v_g)},
\end{equation}
allowing us to eliminate $c(x)$. After some manupulation, we obtain the differential equation
\begin{equation}
\label{eq:DE2}
f_{\rm out}''(x) + a f_{\rm out}'(x) + b f_{\rm out}(x) = f_0''(x) + \tilde{a} f_0'(x) + \tilde b f_0(x).
\end{equation}

We now have two coupled second-order differential equations. By operating on Eq.~\ref{eq:DE2} with $\left[ d^2/dx^2 - a d/dx + b \right]$ we obtain a single differential equation
\begin{equation}
\label{eq:DE2}
\left[ \frac{d^4}{d x^4} + (2b - a^2) \frac{d^2}{dx^2} + b^2 \right] f_{\rm out} (x) = \left[ \frac{d^4}{d x^4} + (2b - a^2) \frac{d^2}{dx^2} + b^2 \right] f_{\rm in} (x) + c \frac{d^2 f_{\rm in}}{dx^2} + d \, f_{\rm in}(x),
\end{equation}
where,
\begin{align}
c &= \frac{2 i \kappa}{v_g} \left[\Delta_{\rm C} - \frac{E}{2} \right]\\
d &= i \frac{E^2 \kappa}{2 v_g^4} \left[ \Delta_{\rm C} - \frac{E}{2} + \frac{2 g^2}{E} \right].
\end{align}
To find the eigenstates we then require $f_{\rm out}(x) = \lambda f_{\rm in}(x)$, where $\lambda$ is a scattering eigenvalue giving the transmission coefficient of the eigenstate. We then have the homogeneous differential equation
\begin{equation}
\label{eq:DE_FINAL}
\left[\frac{d^4}{d x^4} + X \frac{d^2}{d x^2} + Y \right] f_{\rm in} (x)=0,
\end{equation}
where,
\begin{align}
X &= 2b - a^2 -\frac{c}{\lambda - 1},\\
Y &= b^2 - \frac{d}{\lambda - 1}.
\end{align}
Equation \ref{eq:DE_FINAL} has the general solution
\begin{equation}
\label{eq:genSol}
f_{\rm in}(x) = A e^{i \Delta_1 x} + B e^{-i \Delta_1 x} + C e^{i \Delta_2 x} + D e^{- i \Delta_2 x},
\end{equation}
where,
\begin{align}
\label{eq:Delta_1AND2}
\Delta_1 &= \frac{\sqrt{X + \sqrt{X^2 - 4Y}}}{2},\\
\Delta_2 &= \frac{\sqrt{X - \sqrt{X^2 - 4Y}}}{2}.
\end{align}

We previously interpreted the factor in the exponent of the center-of-mass motion as being proportional to the total two-photon energy. Here the factors $\Delta_1$ and $\Delta_2$ in the exponents are proprtional to the difference between the energy of the two photons. Rearranging (\ref{eq:Delta_1AND2}) we can get the scattering eigenvalue $\lambda$ in terms of $\Delta_1$ and $\Delta_2$.  Rearranging both equations in (\ref{eq:Delta_1AND2}) leads to the same functional form for the scattering eigenvalue,
\begin{equation}
\label{eq:lambda}
\lambda(E, \Delta_1) = 1 + \frac{d - c \Delta_1^2}{a^2 \Delta_1^2 + (b - \Delta_1)^2} = \lambda(E, \Delta_2) =  1 + \frac{d - c \Delta_2^2}{a^2 \Delta_2^2 + (b - \Delta_2)^2}.
\end{equation}
This indicates that $\Delta_1$ and $\Delta_2$ are not both independent variables, but that one follows from the other. We will choose to use $\Delta_1$ as the independent variable and define $\Delta_2$ in terms of $\Delta_1$. Solving (\ref{eq:lambda}) for $\Delta_2$ gives
\begin{equation}
\label{eq:Delta2}
\Delta_2 = \pm \frac{\sqrt{b^2 c + a^2 d - 2 b d + d \Delta_1^2}}{\sqrt{c \Delta_1^2 - d}} = \pm \frac{\sqrt{ 4 E g^2 \kappa ^2+\left[E (E-2 \Delta_{\rm C})-4
   g^2\right] \left[E^3+4 g^2 (3 E-2 \Delta_{\rm C})-4 \Delta_1^2
   E  v_g ^2\right] }}{2  v_g  \sqrt{ -4 E g^2-(E-2 \Delta_{\rm C}) (2 \Delta_1  v_g -E) (E+2 \Delta_1
    v_g )}}.
\end{equation}
We note that the scattering eigenvalue in Eq.~\ref{eq:lambda} can be rearranged to a more intuitive form,
\begin{equation}
\lambda(E, \Delta_1) = \frac{\frac{E}{2} + v_g \Delta_1 - \Delta_{\rm C} - \frac{g^2}{E/2 + v_g \Delta_1} - i \frac{\kappa}{2}}{\frac{E}{2} + v_g \Delta_1 - \Delta_{\rm C} - \frac{g^2}{E/2 + v_g \Delta_1} + i \frac{\kappa}{2}} 
\frac{\frac{E}{2} - v_g \Delta_1 - \Delta_{\rm C} - \frac{g^2}{E/2 - v_g \Delta_1} - i \frac{\kappa}{2}}{\frac{E}{2} - v_g \Delta_1 - \Delta_{\rm C} - \frac{g^2}{E/2 - v_g \Delta_1} + i \frac{\kappa}{2}} = t_{E/(2v_g) + \Delta_1} t_{E/(2 v_g) - \Delta_1},
\end{equation}
where $t_k$ is the single-photon transmission coefficient from (\ref{eq:onePhotScattMat}) and $E/2 \pm v_g \Delta_1$ gives the energy of each of the two photons. The two-photon transmission coefficient is therefore a product of two single-photon transmission coefficients for the energy of the two photons.

In order to proceed we need to use the boundary conditions of the problem to derive the values $B/A$, $C/A$, and $D/A$. To do this we first need to compute $f_0(x)$, which we obtain by solving (\ref{eq:DE1}) directly,
\begin{equation}
\begin{split}
f_0(x) = &\frac{\left[ -\Delta_1^2 - i a \Delta_1 + b + i \kappa \Delta_1/v_g + i E \kappa/(2 v_g^2) \right] A e^{i \Delta_1 x}}{b - i a \Delta_1 - \Delta_1^2} + \frac{ \left[ -\Delta_1^2 + i a \Delta_1 + b - i \kappa \Delta_1/v_g + i E \kappa/(2 v_g^2) \right]B e^{-i \Delta_1 x}}{b + i a \Delta_1 - \Delta_1^2}\\
& + \frac{ \left[ -\Delta_2^2 - i a \Delta_2 + b + i \kappa \Delta_2/v_g + i E \kappa/(2 v_g^2) \right]C e^{i \Delta_2 x}}{b - i a \Delta_2 - \Delta_2^2} + \frac{\left[ -\Delta_2^2 + i a \Delta_2 + b - i \kappa \Delta_2/v_g + i E \kappa/(2 v_g^2) \right]D e^{-i \Delta_2 x}}{b + i a \Delta_2 - \Delta_2^2}\\
= & t_{E/(2v_g) + \Delta_1} A e^{i \Delta_1 x} + t_{E/(2v_g) - \Delta_1} B e^{ - i \Delta_1 x} + t_{E/(2v_g) + \Delta_2} C e^{i \Delta_2 x} + t_{E/(2 v_g) - \Delta_2} D e^{-i \Delta_2 x}.
\end{split}
\end{equation}  
In the second line the form of $f_0(x)$ is intuitive: one of the two photons has scattered off the cavity and therefore each term acquires a single transmission coefficient.  

\subsubsection{Boundary conditions}

We now use Eqs.~\ref{eq:alphaEq}-\ref{eq:cEq} to derive boundary conditions for $f_{\rm in}(x)$. Firstly, Eqs.~\ref{eq:c0Eq} and \ref{eq:eEq} are boundary conditions that relate $c_0$ and $c(0)$ as well as $b_0$ and $b(0)$. Again, we use $c(0) = \left[ c(0^+) + c(0^-) \right]/2$ and $b(0) = \left[ b(0^+) + b(0^-) \right]/2$. We obtain two more equations by integrating (\ref{eq:bEq}) and (\ref{eq:cEq}) over the origin,
\begin{align}
e &= \frac{i v_g}{\eta}\left[b(0^+) - b(0^-) \right],\\
c_0 &= \frac{i v_g}{\sqrt{2} \eta} \left[c(0^+) - c(0^-) \right].
\end{align}
Eliminating $c_0$ and $e$ then gives two equations
\begin{align}
(2\Delta_{\rm C} - E)\frac{i v_g}{\sqrt{2} \eta} \left[c(0^+) - c(0^-) \right] + \frac{\eta}{\sqrt{2}} \left[c(0^+) + c(0^-) \right] + \frac{i \sqrt{2}v_g g}{\eta} \left[b(0^+) - b(0^-) \right] &= 0,\\
(\Delta_{\rm C} - E)\frac{i v_g}{\eta} \left[b(0^+) - b(0^-) \right] + \frac{\eta}{2} \left[b(0^+) + b(0^-) \right] + \frac{i v_g g}{\eta} \left[c(0^+) - c(0^-) \right] &= 0.
\end{align}
We now relate these boundary conditions to $f_{\rm in}(x)$ and $f_0(x)$. To do this we use (\ref{eq:c_for_f0_fin}) and (\ref{eq:c_for_fout_f0}) to obtain $c(x)$ in terms of $f_{\rm in}(x)$ and $f_0(x)$ when $x<0$ and $x>0$ respectively. We then use (\ref{eq:bEq}) to obtain $b(x)$ in terms of $c(x)$ and thus in terms of $f_{\rm in}(x)$ and $f_0(x)$ when $x<0$ and $x>0$. Using these we obtain two boundary conditions,
\begin{align}
\label{eq:BC1}
\left. \frac{d f_{\rm in}}{d x} \right|_{x=0} &= 0,\\
\label{eq:BC2}
2 i v_g \left. \frac{d f_0}{dx} \right|_{x=0}  &=  \left[ (\lambda+1) i K_1  + (\lambda-1) K_3 \right] f_{\rm in}(0) + 2 i K_2 f_0(0) ,
\end{align}
where $K_1 = -2(E - \Delta_{\rm C})(E/2 - \Delta_{\rm C})/\kappa +  \kappa/2 + 2  g^2 / \kappa$,  $K_2 = 2(E - \Delta_{\rm C})(E/2 - \Delta_{\rm C})/\kappa +  \kappa/2  - 2 g^2/\kappa$, and $K_3 = 3 E/2 - 2 \Delta_{\rm C}$. In terms of the constants $A$, $B$, $C$, and $D$, these conditions are,
\begin{align*}
i \Delta_1(A - B) + i \Delta_2 (C - D) &= 0,\\
(A + B + C + D)\left[ (\lambda+1) i K_1  + (\lambda-1) K_3 \right] &= -2 \left[ (i K_2 + v_g \Delta_1)t_{E/(2v_g) + \Delta_1}A + (i K_2 - v_g \Delta_1)t_{E/(2v_g) - \Delta_1}B \right.\\
&\left. + (i K_2 + v_g \Delta_2) t_{E/(2 v_g) + \Delta_2} C + (i K_2 - v_g \Delta_2)t_{E/(2 v_g) - \Delta_2} D \right].
\end{align*}
Here we have used a concise notation with $\lambda = \lambda(E, \Delta)$. These two boundary conditions, along with orthonormality, can be used find a class of extended states. We do not use the extended states in the main text for any of the computations.

\subsubsection{Bound States}

The general form of the two-photon eigenstates solution (\ref{eq:genSol}) is composed of a linear combination of four exponential functions. When $\Delta_1$ and $\Delta_2$ are real, these lead to plane-wave or spatially extended solutions. We are interested in bound-state solutions where the photons are exponentially localised in about the difference coordinate, i.e., solutions where $\Delta_1$ and $\Delta_2$ are imaginary or complex with a positive-values imaginary part for $x>0$. Here $\Delta_1$ is no longer a quantum number labelling the state meaning the state is only labelled by its total two-photon energy $E$. For these solutions to remain bounded (not go to infinite), we then require $B=0$ and $D=0$. The two boundary conditions (\ref{eq:BC1}) and (\ref{eq:BC2}) give two coupled equations which can be written in matrix form as $\mathbf M \mathbf v = \mathbf 0$. Here, $\mathbf v = \begin{pmatrix}  A & C \end{pmatrix}^T$. We obtain a condition to find $\Delta_1$ by requiring the determinant of the matrix vanish $\operatorname{det(\mathbf M)} = 0$, 
\begin{equation}
\label{eq:detM}
\frac{2 K_2 (t_{\Delta_1} \Delta_2 - t_{\Delta_2} \Delta_1) + 2 i v_g \Delta_1 \Delta_2 (t_{\Delta_2} - t_{\Delta_1}) + i K_3 (\Delta_1 - \Delta_2) (\lambda - 1) - K_1 (\Delta_1 - \Delta_2) (1 + \lambda)}{2 K_2 (t_{-\Delta_2} \Delta_1 - t_{-\Delta_1} \Delta_2) + 2 i v_g \Delta_1 \Delta_2 (t_{-\Delta_2}  - t_{-\Delta_1})  - i K_3 (\Delta_1 - \Delta_2) (\lambda - 1) + K_1 (\Delta_1 - \Delta_2) (1 + \lambda)} = 0
\end{equation}
Here, we have used the concise notation $t_{\pm \Delta} \equiv t_{E/(2v_g) \pm \Delta}$. The eigenvector with coefficients $A$ and $C$ then corresponds to the eigenvector with zero eigenvalue. In general, the value of $\Delta_1$ that satisfies $(\ref{eq:detM})$ is complex. Furthermore the value of $\Delta_2$ corresponding to this value of $\Delta_1$ is also in general complex. The transmission coefficient for the bound state is then obtained by substituting the complex value of $\Delta_1$ in (\ref{eq:lambda}). Similarly $\Delta_2$ is also obtained by substituting the complex value of $\Delta_1$ in (\ref{eq:Delta2}). We solve these equations numerically and use a two-dimensional minimization algorithm to find the real and imaginary parts of $\Delta_1$.

The general form of the eigenstate is then
\begin{equation}
\begin{split}
\label{eq:boundStateFullDefinition}
|B_E \rangle  &= \frac{1}{\sqrt{2}} \int dx_1 dx_2 \hat{a}^\dagger(x_1) \hat{a}^\dagger(x_2) | 0 \rangle B_E(x_c, x)\\
B_E(x_c, x) &= e^{i E x_c/v_g} N \left\{ A e^{- \operatorname{Im}{(\Delta_1)}|x|} \left[ \cos{(\operatorname{Re}{(\Delta_1)} x)} + i \operatorname{sgn}{(x)} \sin{(\operatorname{Re}{(\Delta_1)} x)} \right] \right.\\
& \left. + C e^{- \operatorname{Im}{(\Delta_2)}|x|} \left[ \cos{(\operatorname{Re}{(\Delta_2)} x)} + i \operatorname{sgn}{(x)} \sin{(\operatorname{Re}{(\Delta_2)} x)} \right]
 \right\} \equiv e^{i E x_c/v_g} B_E(x),
\end{split}
\end{equation}
where
\begin{equation}
N = \frac{1}{\sqrt{2 \pi}} \left\{\frac{|A|^2}{\operatorname{Im}{(\Delta_1)}} + \frac{|C|^2}{\operatorname{Im}{(\Delta_2)}} + 4 \operatorname{Re}{\left[ \frac{A \, C^*}{\operatorname{Im}{(\Delta_1)} + \operatorname{Im}{(\Delta_2)} - i (\operatorname{Re}{(\Delta_1)} - \operatorname{Re}{(\Delta_2)})} \right]}\right\}^{-1/2}.
\end{equation}
Again, we note that the full expression for $B_E(x)$ is obtained by solving for $A$ and $C$ numerically.

\subsubsection{Perturbation approximation}

While the full expression for the bound states can only be obtained numerically, we can perform a perturbative expansion to obtain approximate expressions. We begin by expanding (\ref{eq:detM}) to first order and in $1/\kappa$ and solving for $\Delta_1$. By $1/\kappa$ we refer to any term where the numerator has units of frequency and the denominator has the value $\kappa$. This gives
\begin{equation}
\Delta_1 = \frac{i \Gamma}{2 v_g} + \frac{i \Gamma^2}{2 v_g \kappa} + O\left(\frac{1}{\kappa^2} \right),
\end{equation}
and,
\begin{equation}
C = - A \frac{\Gamma}{\kappa} + O\left(\frac{1}{\kappa^2} \right).
\end{equation}
Substituting this into the expression for (\ref{eq:Delta2}) gives,
\begin{equation}
\Delta_2 = \frac{i \kappa}{2 v_g} + i \frac{E^2 - 2\Gamma^2}{4 v_g \Gamma} + O\left(\frac{1}{\kappa^2} \right).
\end{equation}

Combining these together and computing the normalization gives,
\begin{equation}
B_{E}(x_c, x) = \frac{e^{i \frac{E}{v_g} x_c }}{\sqrt{4 \pi v_g}} \frac{1}{\sqrt{\frac{\kappa}{\Gamma^2 + \Gamma \kappa} + \frac{8 \Gamma^2}{2\Gamma^3 + E^2 \kappa + 2 \Gamma \kappa^2} + \frac{2 \Gamma^3}{\kappa^2 [E^2 + 2 \Gamma (\kappa - \Gamma)]}}} \left[e^{-\frac{\Gamma}{2 v_g} (1 + \frac \Gamma \kappa)|x|}  - \frac{\Gamma}{\kappa} e^{-\frac{\kappa}{2 v_g} \left(1 + \frac{E^2 - 2 \Gamma^2}{2 \kappa \Gamma} \right) |x|} \right] + O\left(\frac{1}{\kappa^2} \right).
\end{equation}
In comparison to the two-photon bound state in waveguide QED \cite{Mahmoodian2020PRX}, the addition of a cavity therefore introduces a new term to the bound state. This term, rather than having a decay length inversely proportional to $\Gamma$ has a decay length inversely proportional to the cavity decay rate $\kappa$. It also has the opposite sign to the other term and therefore the two terms interfere destructively. This causes the cusp at the origin to smooth out for infinitesimally small $\kappa$.

Using the perturbative solution of $\Delta_1$, we obtain for the transmission coefficient to leading order
\begin{equation}
t_{B}(E) = \frac{-2 \Gamma^2 + E (i \kappa - 2 E) + 2\Gamma (\kappa + i E)}{2 \Gamma^2 + E (i \kappa + 2 E) - 2\Gamma (\kappa - i E)} + O\left(\frac{1}{\kappa^2} \right).
\end{equation}
The delay incurred by a resonant pulse of bound states in the center-of-mass coordinate is then
\begin{equation}
\phi'_B(0) =  \frac{1}{\Gamma} + \frac{3}{\kappa} + O\left(\frac{1}{\kappa^2} \right),
\end{equation}
where $\phi_B(E) = -i \log{t_{B}(E)}$. The distortion on resonance is then
\begin{equation}
\phi'''_B(0) = -\frac{1}{2 \Gamma^3} \left[ 1 -  \frac{3 \Gamma}{\kappa} \right] + O\left(\frac{1}{\kappa^2}  \right).
\end{equation}

\subsubsection{Two-photon bound-state scattering, delay, and dispersive effects}

To compute the scattered two-photon state in the bound subsector we simply compute
\begin{equation}
\label{eq:boundOut}
|\textrm{out}\rangle_{\rm bound} = \int dE \,  t_2(E) \, | B_E \rangle \langle B_E| \textrm{in} \rangle, 
\end{equation}
where the bound eigenstates $|B_E \rangle$ are given in (\ref{eq:boundStateFullDefinition}). Here, the input state is taken to be a separable state,
\begin{equation}
|\textrm{in} \rangle = \frac{1}{\sqrt{2}}\int dx_1 dx_2 \hat{a}^\dagger(x_1) \hat{a}^\dagger(x_2) | 0 \rangle f(x_1) f(x_2),
\end{equation}
where, as for single photon scattering, the functions $f(x)$ can be the pulse shape given by the laser (which are approximately Gaussian) or the exact Gaussian functions in (\ref{eq:inputPulseShape}). We compute the output state by calculating the integral over $E$ numerically.

We now show that the delay and the dispersion can be computed in the center-of-mass coordinate. We note that the input state can be written in center-of-mass and difference coordinate, which for a Gaussian input becomes
\begin{equation}
|\textrm{in} \rangle = \frac{1}{\sqrt{2}}\int dx_1 dx_2 \hat{a}^\dagger(x_1) \hat{a}^\dagger(x_2) | 0 \rangle \frac{1}{\sigma \sqrt{\pi}} e^{2i k_0 x_c}e^{-x_c^2/\sigma^2} e^{-x^2/(4\sigma^2)}.
\end{equation}
The overlap of the input state and the bound state $\langle B_E | \textrm{in} \rangle = \nu e^{-(E - 2k_0)^2\sigma^2/4}$, where the constant $\nu$ is the result of the difference-coordinate integral and it's form is not relevant here. The output state can thus be written 
\begin{equation}
| \textrm{out} \rangle_{\rm bound} = \frac{1}{\sqrt{2}} \int dx_1 dx_2 \hat{a}^\dagger(x_1) \hat{a}^\dagger(x_2) | 0 \rangle \nu B_E(x) \int dE e^{i E x_c} t_B(E) e^{-(E - 2k_0)^2\sigma^2/4}. 
\end{equation}  
Here the bound-state transmission coefficient in the center-of-mass coordinate plays the same role as the single-photon transmission coefficient in (\ref{eq:singleOut}). We can therefore find the delay and dispersion in the same way as for the single-photon case by defining
\begin{equation}
\label{eq:boundPhi}
\phi_B(k_0) = \left. -i \log{(t_B(E))}\right|_{E=2 v_g k_0}
\end{equation}
The delay and higher-order dispersion for the two-photon bound state are calculated numerically, by taking first, second and third, derivatives of (\ref{eq:boundPhi}) in $E$.

\subsubsection{Extended States}

The extended scattering eigenstates are those with the form (\ref{eq:genSol}) where the exponents are imaginary and the field is extended in the relative coordinate $x$. Here there are two sets of exponents, two with $\pm \Delta_1$ and two with $\pm \Delta_2$. As discussed, we take $\Delta_1$ to be the quantum number of the eigenstate and $\Delta_2$ then depends on $\Delta_1$ according to (\ref{eq:Delta2}). Depending on the values of $\Delta_1$, $E$, $\kappa$, and $g$, $\Delta_2$ can either be real or imaginary. Noting that in (\ref{eq:Delta2}) we are free to define the sign of $\Delta_2$, we always choose this such that when it is imaginary it has a positive imaginary part. In this case, for the solution to remain bounded, we require $D=0$. We then have two boundary conditions and the requirement that the eigenstate be normalized, which gives three conditions in total.  Using the boundary conditions to solve for $B$ and $C$ gives
\begin{align}
\frac{B}{A} = \frac{(\lambda +1) K_1 (\Delta_1-\Delta_2)-2 \Delta_2 K_2 t_{\Delta_1}+2 \Delta_1 K_2 t_{\Delta_2}-i (\lambda -1) K_3 (\Delta_1-\Delta_2)+2 i \Delta_1 \text{$\Delta$2} t_{\Delta_1}  v_g -2 i \Delta_1 \Delta_2 t_{\Delta_2}  v_g }{(\lambda +1) K_1 (\Delta_1+\Delta_2)+2 K_2 (\Delta_2 t_{-\Delta_1}+\Delta_1 t_{\Delta_2})-i (\lambda -1) K_3
   (\Delta_1+\Delta_2)+2 i \Delta_1 \Delta_2  v_g  (t_{-\Delta_1}-t_{\Delta_2})},\\
\frac{C}{A} = -\frac{2 \Delta_1 ((\lambda +1) K_1+K_2 (t_{\Delta_1}+t_{-\Delta_1})-i (\lambda -1) K_3-i \Delta_1  v_g  (t_{\Delta_1}-t_{-\Delta_1}))}{(\lambda +1) K_1 (\Delta_1+\Delta_2)+2 K_2
   (\Delta_2 t_{-\Delta_1}+\Delta_1 t_{\Delta_2})-i (\lambda -1) K_3 (\Delta_1+\Delta_2)+2 i \Delta_1 \Delta_2  v_g  (t_{-\Delta_1}-t_{\Delta_2})},
\end{align}
where the compressed notation $t_\Delta \equiv t(E/2 + v_g \Delta)$ has been used.

In addition to begin imaginary, $\Delta_2$ can also be real. Here, in general, the constant $D \neq 0$. We therefore have four unknown parameters and only two boundary conditions and one normalization condition. The lack of a boundary condition arises because the system has a degeneracy. In (\ref{eq:genSol}) one can interchange $\Delta_1$ and $\Delta_2$ and the solution remains valid and has the same eigenvalue (\ref{eq:lambda}). In order to obtain a further constraint we can require that the two degenerate solutions,
\begin{align}
f_{\rm in}^{\Delta_1}(x) = A_{\Delta_1} e^{i \Delta_1 x} + B_{\Delta_1} e^{-i \Delta_1 x} + C_{\Delta_1} e^{i \Delta_2 x} + D_{\Delta_1} e^{-i \Delta_2 x},\\
f_{\rm in}^{\Delta_2}(x) = A_{\Delta_2} e^{i \Delta_2 x} + B_{\Delta_2} e^{-i \Delta_2 x} + C_{\Delta_2} e^{i \Delta_1 x} + D_{\Delta_2} e^{-i \Delta_1 x},
\end{align}
are orthogonal. Here we have used the fact that the function $\Delta_2 = h(\Delta_1)$ in (\ref{eq:Delta2}) is its own inverse, i.e., $\Delta_1 = h(\Delta_2)$. The orthogonality condition can be achieved by requiring $\Delta_1>0$, $\Delta_2>0$, and 
\begin{align}
\frac{B_{\Delta_1}}{A_{\Delta_1}} &= - \frac{C_{\Delta_2}^*}{D_{\Delta_2}^*},\\
\frac{B_{\Delta_2}}{A_{\Delta_2}} &= - \frac{C_{\Delta_1}^*}{D_{\Delta_1}^*}.
\end{align}
This then gives a sufficient number of constraints to find all unknowns.

\bibliographystyle{naturemag_noURL}
\bibliography{WignerDelay_SI.bib}